\documentclass[
reprint,
superscriptaddress,
 amsmath,amssymb,
 aps,
]{revtex4-2}

\usepackage{graphicx}
\usepackage{dcolumn}
\usepackage{textcomp}
\usepackage{bbold}
\usepackage{bm}
\usepackage{gensymb}

\usepackage{siunitx}
\usepackage[linktocpage,colorlinks=true,linkcolor=blue,citecolor=blue,breaklinks=true,urlcolor=blue]{hyperref}
\usepackage[T1]{fontenc}
\usepackage{xcolor}
\usepackage{graphicx}  
\usepackage{dcolumn}   
\usepackage{amssymb,amsmath,bm}  
\usepackage[bottom]{footmisc}
\usepackage{physics}
\usepackage{siunitx}
\usepackage{titlesec}

\titlespacing\section{92pt plus 2pt minus 92pt}{12pt plus 4pt minus 2pt}{4pt plus 2pt minus 2pt}
\titlespacing\subsection{0pt}{12pt plus 4pt minus 2pt}{4pt plus 2pt minus 2pt}
\titlespacing\subsubsection{0pt}{12pt plus 4pt minus 2pt}{2pt plus 1pt minus 1pt}

\titleformat{\section}
   { \normalfont\large\bfseries}{\thesection}{0.4 em}{\centering  }

\begin{document}

\title{Enhancement of bacterial rheotaxis in non-Newtonian fluids}

\author{Bryan O. Torres Maldonado}
\thanks{B.T. \& A.T. contributed equally to this work and are joint first authors}
 \affiliation{%
Department of Mechanical Engineering \& Applied Mechanics,
		University of Pennsylvania, Philadelphia, PA 19104
}%
\author{Albane Th{é}ry}
\thanks{B.T. \& A.T. contributed equally to this work and are joint first authors}
 \affiliation{%
Department of Mathematics,
		University of Pennsylvania, Philadelphia, PA 19104
}
\author{Ran Tao}
\affiliation{{Department of Physics \& Astronomy,
		University of Pennsylvania, Philadelphia, PA 19104}}
\author{Quentin Brosseau}
\affiliation{%
Department of Mechanical Engineering \& Applied Mechanics,
		University of Pennsylvania, Philadelphia, PA 19104
}%
\author{Arnold J. T. M. Mathijssen}
 \email{amaths@sas.upenn.edu}
\affiliation{{Department of Physics \& Astronomy,
		University of Pennsylvania, Philadelphia, PA 19104}}
\author{Paulo E. Arratia}%
 \email{parratia@seas.upenn.edu}
 \affiliation{%
Department of Mechanical Engineering \& Applied Mechanics,
		University of Pennsylvania, Philadelphia, PA 19104
}


\begin{abstract}
Bacteria often exhibit upstream swimming, which can cause the contamination of biomedical devices and the infection of organs including the urethra or lungs. This process, called rheotaxis, has been studied extensively in Newtonian fluids. However, most microorganisms thrive in non-Newtonian fluids that contain suspended polymers such as mucus and biofilms. Here, we investigate the rheotatic behavior of \textit{E. coli} near walls in non-Newtonian fluids. Our experiments demonstrate that bacterial upstream swimming is enhanced by an order of magnitude in shear-thinning polymeric fluids relative to Newtonian fluids. This result is explained by direct numerical simulations, revealing a torque that promotes the alignment of bacteria against the flow. From this analysis, we develop a theoretical model that accurately describes experimental rheotatic data in both Newtonian and shear-thinning fluids.
\end{abstract}

\date{\today}

\maketitle
\onecolumngrid
\vspace*{12pt}
\twocolumngrid

Many microswimmers are immersed in external flows, like circulatory flows or oceanic turbulence, that significantly affect their swimming behavior \cite{vogel2020life, Wheeler_2019}. 
The coupling between external, self-generated active flows leads to unusual transport properties \cite{Kessler_1985, ran_pnas_2021}. 
In particular, many microswimmers display rheotaxis, which is the ability to reorient against flows and move upstream \cite{Kaya_2012, Marcos_2012, Kantsler_2014,brosseau2019relating, Mathijssen_2016, Mathijssen_2019} through a mechanical interplay between activity and a guiding external shear flow. 
Rheotaxis has been observed in sperm cells \cite{Bretherton1961}, ciliates \cite{Takuya_cilia_2021}, flagellated \cite{Mathijssen_2019, Hill_2007, Kaya_2012,Figueroa_Morales_2020} and non-flagellated bacteria \cite{Meng_2005}, and microrobots \cite{ren2017rheotaxis, brosseau2019relating}. 
For bacteria, the ability to resist and overcome adverse flows is a selective advantage for pathogens colonizing environments where liquids are flowing, including the urinary tract, bloodstream, or hospital tubing \cite{Siryaporn_2015, Figueroa_Morales_2020}.  

Bacterial rheotaxis occurs in bulk flows \cite{Jing_2020, Ronteix2022} but upstream swimming is significantly stronger in the vicinity of walls \cite{Hill_2007, Kaya_2012, Mathijssen_2019, Figueroa_Morales_2020}. 
The mechanical response to fluid flow gradients near surfaces results from the interactions of the swimmer with the flow and the wall. 
Bacteria swimming near a surface experience alignment due to steric and hydrodynamic interactions \cite{berke2008hydrodynamic,  Spagnolie_2012},
and are directed towards the wall in clockwise circular trajectories due to hydrodynamic torque \cite{Spagnolie_2012, Lauga_2006}. The shear flow produces a ``weathervane effect'' that aligns the flagellum downstream, allowing the swimmer to reorient against the flow \cite{Mathijssen_2019, Kantsler_2014,tung2015emergence}.
Together, the combined effects of the wall, shear flow, and self-propulsion enable bacteria to swim upstream \cite{Kaya_2012, Mathijssen_2019}.

Much of our current understanding of rheotaxis derives from Newtonian fluids \cite{Kaya_2012,zottl_periodic_2013}. Many microorganisms, however, inhabit natural environments (from soil to the human body) that contain suspended particles and/or (bio)polymers and often exhibit non-Newtonian behavior such as shear-thinning viscosity behavior and viscoelasticity \cite{ Fauci2006biofluid, Arratia_PhysRevFluids2022, Spagnolie_ARFM_2023}. 
The non-Newtonian properties of the surrounding fluid influence every aspect of microscale propulsion \cite{Arratia_PhysRevFluids2022, Spagnolie_ARFM_2023}: 
depending on propulsion mode and fluid properties, it can lead to an increase \cite{Li2015undulatory, Gomez_2016, Demir2020nonlocal, Kamdar_2022} 
or a decrease \cite{Shen2011undulatory, Datt_2015} in swimming speed and affects the interaction of swimmers with external flows \cite{ Li_2015} and boundaries \cite{ Cao_2022}. 
Investigating microorganisms swimming upstream in model shear-thinning fluids, for example, can provide crucial insight into the pathways of pathogens in the body \cite{Schreiber_2004}, 
fertility \cite{Fauci2006biofluid}, 
and food safety. Thus, locomotion in complex systems and its interplay with external cues have attracted significant experimental and theoretical attention in recent years \cite{Spagnolie_ARFM_2023}. 
However, despite its biomedical significance, little is known about rheotaxis in complex fluids. It has been investigated using theory and numerical simulations \cite{mathijssen2016hydrodynamics}, but experiments exploring the relationship between non-Newtonian fluid properties and microbial upstream swimming remain scarce. Here, we aim to address this gap in the literature. The current manuscript focuses predominately on shear-thinning viscosity effects.

We combine experiments, numerical simulations, and theory to investigate the upstream motion of \textit{E. coli} in Newtonian and polymeric fluids at varying flow rates near a surface. Our experiments show that shear-thinning viscosity behavior can result in a significant enhancement of upstream swimming, even for dilute solutions. Bacterial trajectories are rectified due to shear-thinning effects near walls, which explains the increase in upstream swimming.  We develop a simple model for the bacterial orientation and position that captures their average swimming speed parallel and perpendicular to the imposed flow.

 \begin{figure*}
 \centering
 \includegraphics[width = 1.9\columnwidth]{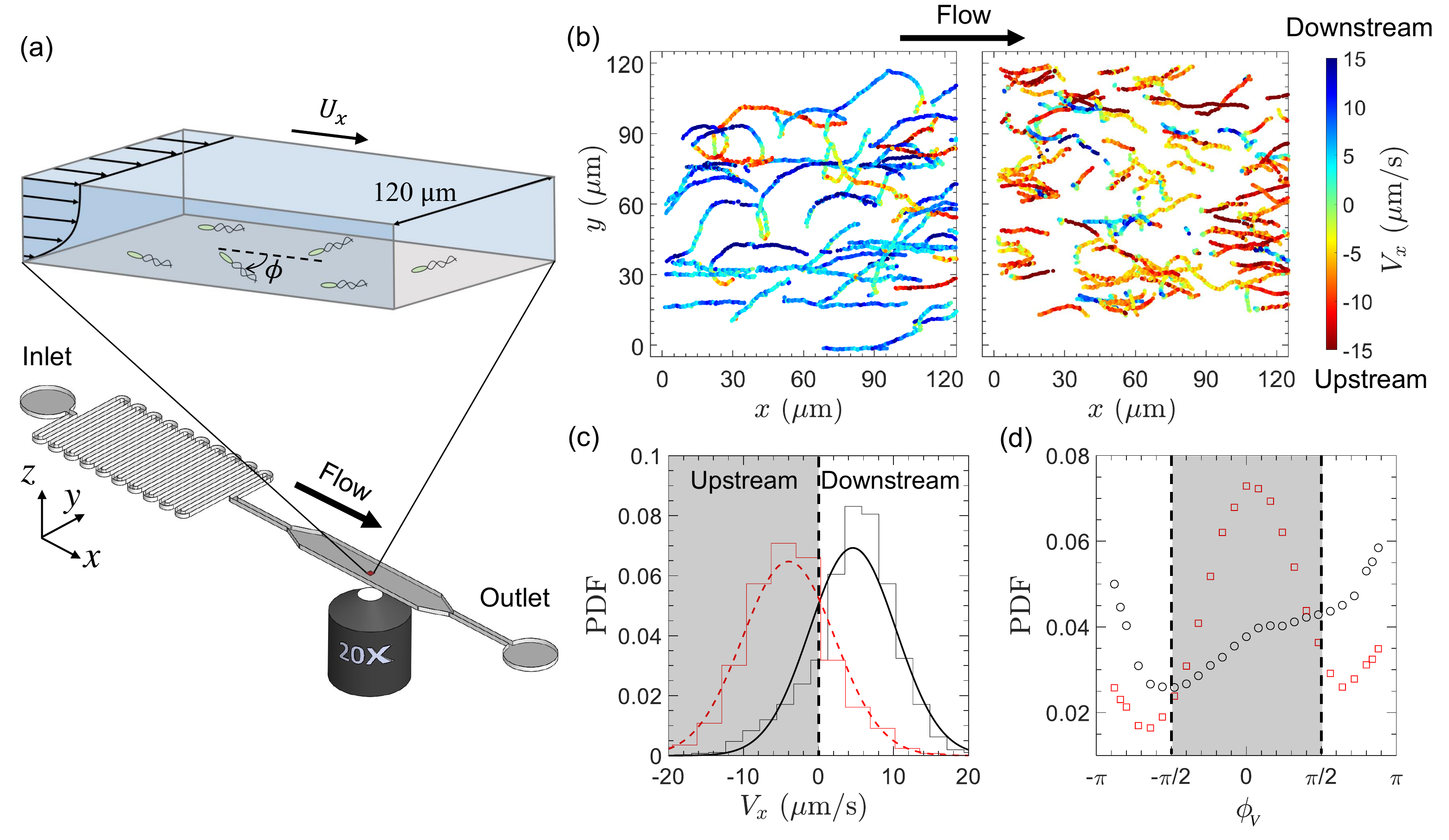} 
 \caption{Experimental set-up and enhanced rheotactic behavior in a shear-thinning fluid compared to a Newtonian one at a given flow rate.  (a) Microfluidic channel, with a long delay line to minimize fluctuations. The region of interest ($ 120\,\mu \textrm{m}\times 120\,\mu \textrm{m}$) is located near the center of the wide part of the channel, away from the vertical walls, and close to the bottom wall. 
 (b) Trajectories of \textit{E. coli} in water (left) and $c/c^*=0.6$ of XG (right), with an applied external flow rate of $20\,\mu $L/hr (top view). The tracks of \textit{E. coli} in XG show a bias in the upstream direction, whereas those in water display fewer swimmers going against the flow. These results suggest that the shear-thinning properties of the fluid contribute to bacterial rheotaxis. 
 (c) PDFs of bacterial swimming velocity parallel to the flow direction $V_x$ for both water (black) and $c/c^*=0.6$ of XG (red) under an external flow rate of $20\,\mu $L/hr. (d) PDFs of the orientation of the bacteria trajectory relative to the flow direction $\phi_v$ for both water (black) and $c/c^*=0.6$ of XG (red) under the same external flow rate $ Q = 20\,\mu $L/hr. Note that $\phi_V$ is different from the angle $\phi$ of the bacterium itself with the flow direction because the swimmer also gets advected.  }
 \label{Fig_1}
\end{figure*}

\section{Experiments with non-Newtonian fluids}
We track \textit{E. coli} near a surface in the presence of an external flow, and measure the velocity of the bacteria against the flow and perpendicular to it. We are interested in the effect of the surrounding fluid being shear-thinning (ST).  Experiments are performed in a microfluidic channel with a controlled external volumetric flow, $Q$ (see Fig.~\ref{Fig_1}a and methods). The region of interest is positioned on the bottom wall in the fully developed flow region, away from the channel's corners to avoid shear-trapping and edge currents effects \cite{Rusconi_2014, Figueroa_Morales_2015}. We use both Newtonian and shear-thinning fluids in the channel. For Newtonian fluids, we perform experiments with buffer and Ficoll solutions to control for the effect of enhanced Newtonian viscosity; these solutions provide a viscosity range from 1 cp to 8 cp. Aqueous solutions of xanthan gum (XG), a semi-rigid polymer, are used to produce shear-thinning fluids; polymer concentrations range from $0.2c^*$ to $c^*$, where $c^*$ denotes the overlap concentration (see SM). We also experimented with (dilute) solutions of a more flexible polymer, namely carboxymethyl cellulose (CMC), that produces fluids with a shear-thinning viscosity and some elasticity \cite{Shen_2012, patteson_running_2015}. All fluids are characterized using a stress-controlled rheometer, and the shear rates ($\dot{\gamma}$) near the surface of the microfluidic channel are obtained using numerical simulations (see Supplementary Material). 

\begin{figure}
 \centering
\includegraphics[width=.9\columnwidth]{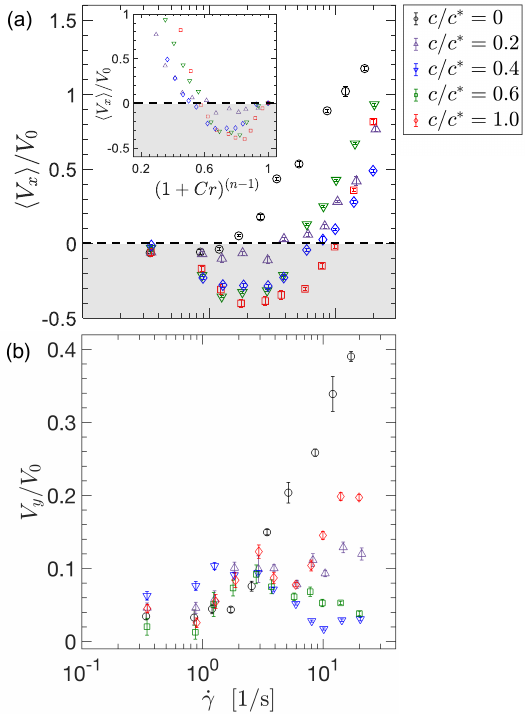}
 \caption{Average swimming velocity of \textit{E. coli} parallel and perpendicular to the flow direction. 
  (a) Average swimming velocity parallel to the flow direction, $V_x/V_0$, where $V_0$ is the mean swimming speed without an external flow ($Q=0\,\mu$L/hr) as a function of shear rate $\dot{\gamma}$. Results are shown for water (black) and XG solutions at $c/c^*=0.2$ (violet), $c/c^*=0.4$ (green), $c/c^*=0.6$ (red), $c/c^*=1.0$ (blue). The data reveals a substantial enhancement of upstream swimming in XG solutions, which also occurs over a broader range of flow rates. The inset graph shows $V_x/V_0$ as a function of a normalized shear rate $(1+Cr)^{n-1}$, where Cr is the Carreau number. (b) Average swimming velocity perpendicular to the flow direction, $V_y/V_0$, as a function of shear rate, showing a reduction in the net lateral drift $V_y$ in polymer solutions. 
  }
 \label{Fig_2}
\end{figure}

Swimming \textit{E. coli} cells are tracked near the bottom wall of the microfluidic chamber \cite{patteson_running_2015}, and we extract their average swimming speed in the parallel $V_x$ and perpendicular $V_y$ directions to the flow (Fig.~\ref{Fig_2}). 

\subsection{No-flow Condition:} In the absence of imposed flow, bacteria display the expected clockwise circular trajectories \cite{Spagnolie_2012, Lauga_2006}. These circular trajectories become less defined, noisier, and display larger mean curvatures in shear-thinning fluids (see SI). This suggests that shear-thinning viscosity behavior affects the swimmer-wall interaction~\cite{Cao_2022, Chen_2021}. 

Next, we measure the mean swimming speed $V_0$ of the motile bacterial population under quiescent conditions. In Newtonian fluids (Ficoll), the \textit{E. coli} mean swimming speed decreases monotonically as viscosity increases. For dilute shear-thinning fluids, on the other hand, the mean swimming speed of \textit{E. coli} increases with polymer concentration, even as the steady fluid viscosity is also increasing (see SI, Fig.2). We observe speed enhancement of \textit{E. coli} near walls of up to $50 \%$ compared to Newtonian fluids, consistent with previous studies in the bulk \cite{patteson_running_2015, Martinez_2014,kamdar_colloidal_2022}. 

\subsection{Applied Flow Condition: Rheotaxis} Typical \textit{E. coli} trajectories under an applied flow are shown in Fig.~\ref{Fig_1}b for $Q=20\,\mu$L/hr in Newtonian (water) and shear-thinning (XG $c/c^*=0.6$) fluids. While most bacteria in the Newtonian fluid (left panel) are dragged downstream (positive $x$-direction) by the flow, there are significant bacteria upstream motion in the ST fluid (right panel).  
We quantify these observations using the probability density functions (PDFs) of cell swimming velocity in the axial direction $V_x$ for Newtonian and ST fluids under the same flow rate in Fig.~\ref{Fig_1}c. The Newtonian speed distribution is centered at (positive) approximately $5 \mu$m/s, showing a downstream bias, while the shear-thinning distribution is centered at approximately - $5 \mu$m/s with an upstream bias. The ST distribution is also broader than the Newtonian ones. These trends are consistent with the representative trajectories shown in Fig.~\ref{Fig_1}b. The PDFs of \textit{E. coli} orientation relative to the flow direction in Fig.~\ref{Fig_1}d confirm that bacteria in ST fluids show a pronounced alignment with the flow upstream direction (relative to the Newtonian case). This is not a purely viscous effect, since bacteria are further advected downstream in a Newtonian medium with higher viscosity $\mu \geq 2\,$mPa$\,$s$^{-1}$ (SI, Fig.~S4). Overall, these results highlight the significant impact of fluid shear-thinning properties on the rheotatic behavior of microorganisms. Shear-thinning viscosity behavior enables positive rheotaxis at flow rates beyond the accessible range in Newtonian fluids.

We now investigate bacterial rheotatic behavior near walls as a function of local flow shear rate ($\dot{\gamma}$) for Newtonian fluids of different viscosities and shear-thinning fluids of different concentrations. Figure~\ref{Fig_2}a shows the normalized average swimming speed parallel to the flow direction ($V_x/V_0$) as a function of $\dot{\gamma}$, where $V_0$ is the bacteria mean speed under quiescent conditions.  In a Newtonian fluid (buffer solution), we observe limited rheotaxis at low shear rates $\dot{\gamma}\approx 0.3-2.4\,$s$^{-1}$ (Fig.~\ref{Fig_2}a dark circles); as $\dot{\gamma}$ increases, nearly the entire bacteria population gets advected downstream. Notably, any bacterial rheotatic behavior is lost in fluids with higher Newtonian viscosities (Ficoll) (see SI Fig.~S4).  However, for shear-thinning fluids (XG solutions), we find (i) more pronounced rheotatic behavior (relative to the Newtonian case) and (ii) rheotaxis over a broader range of $\dot\gamma$ (from 0.9 to 9 s$^{-1}$), as shown in Fig.~\ref{Fig_2}a (color non-circle symbols). At sufficiently high shear rates, all bacteria are dragged downstream in both Newtonian and ST cases.  Note that similar behavior is found with CMC solutions (SI, Fig.~S5), although some differences are found due to the flexible nature of the CMC molecule, which leads to fluid elasticity.

Next, we examine the average swimming speed of bacteria perpendicular to the flow direction ($V_y$ ) as a function of shear rates $\dot{\gamma}$ (Fig.~\ref{Fig_2}b). For the Newtonian case, bacteria are known to swim toward the positive $y$-axis as a result of their clockwise trajectories at the wall in quiescent fluids \cite{Lauga_2006, Mathijssen_2019}. This tendency increases with shear rate as bacteria transition from loopy to oriented trajectories \cite{Mathijssen_2019}. For shear-thinning fluids, on the other hand, we find a relative decrease in $V_y$ for all XG solutions. Combined with the $V_x$ data (Fig.~\ref{Fig_2}a), this observation indicates that shear-thinning viscosity can significantly enhance bacteria alignment with the flow.

Our experimental results show that only a small portion of the population can swim upstream in Newtonian fluids, while the majority is advected downstream even at very low shear rates. However, in shear-thinning fluids, a large fraction of the same bacteria population can swim upstream up to the point where the local flow velocity becomes comparable to their swimming speed. That is, bacterial rheotaxis is significantly enhanced (relative to Newtonian fluids) in ST fluids. What are the main mechanisms governing this behavior?

\section{Model, mechanisms, \& discussion}
In this section, we develop a model to describe the experimental observation and gain insight into the main mechanism responsible for the enhancement in rheotactic behavior in shear-thinning fluids near walls. We focus our efforts on the concentration that leads to the strongest level of rheotaxis, $c = 0.6 c^*$. 

\begin{figure}[t]
 \centering
 \includegraphics[width= \columnwidth]{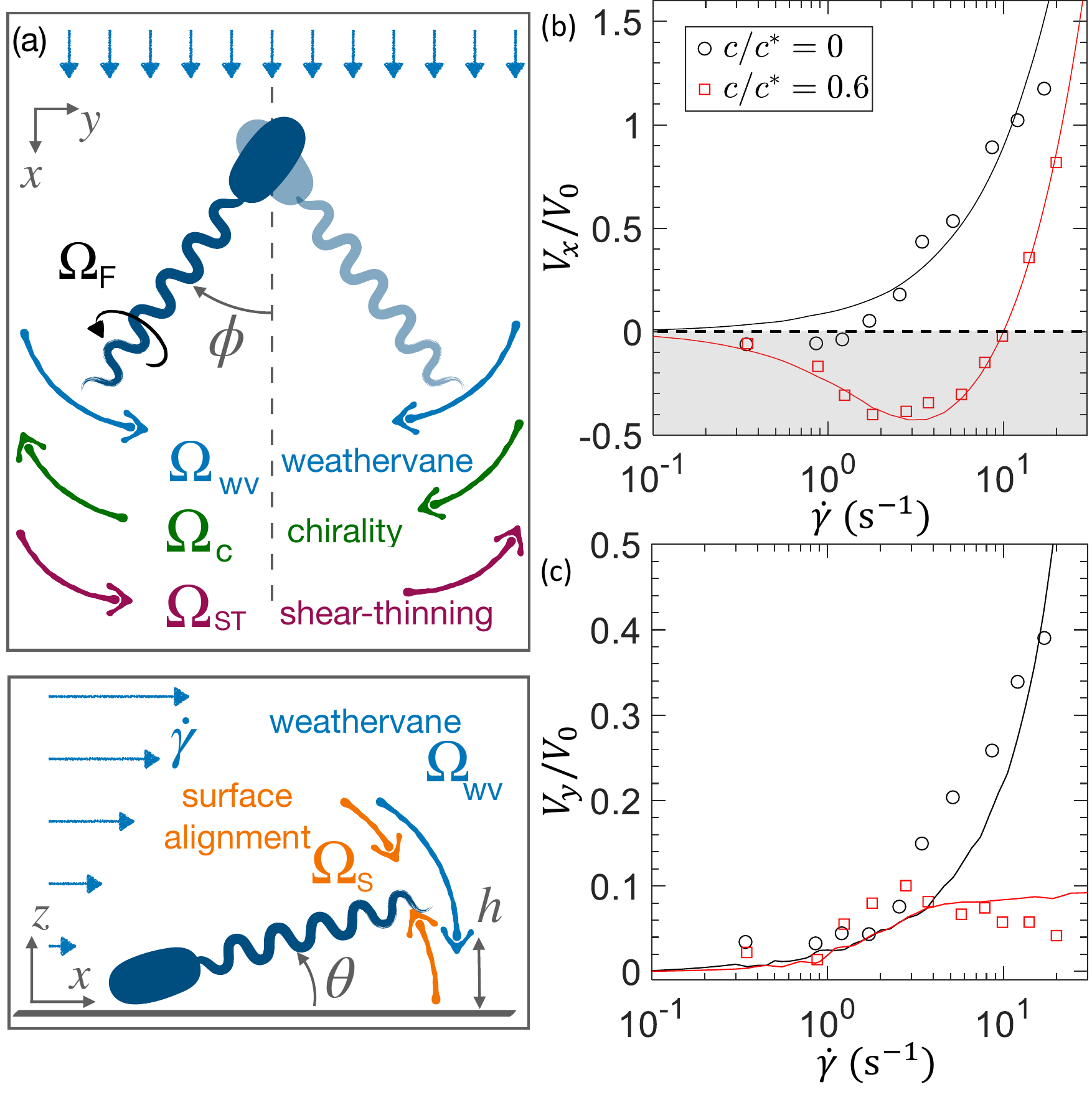}
 \caption{ A minimal model for the orientation of the bacteria captures both the enhanced rheotaxis and the decreased lateral drift in a shear-thinning fluid. (a) Reorientation mechanisms for a bacterium at the wall, as listed in the model. In the horizontal plane (top), the weathervane effect causes the bacterium to reorient towards the upstream direction. The chirality of the bacterium's flagella induces a drift towards the negative y-direction. However, this effect is counteracted in shear-thinning fluids by the rotation-induced drift. In the vertical direction (bottom), the bacterium tends to swim at an angle in a quiescent fluid. Meanwhile, the weathervane effect encourages horizontal alignment with the flow. We perform a simulation of $50,000$ swimmers and calculate their average swimming velocities. (b) Average swimming velocity of \textit{E.coli} parallel to the flow direction $V_x$ in Newtonian (water) and a shear-thinning fluid (XG), normalized by $V_0$, as a function of shear rate $\dot \gamma$. Experimental measurements for water and XG at $c/c^*=0.6$ are shown as black circles and red squares, respectively. Solid lines represent simulations of each fluid, revealing a notable increase in bacterial swimming upstream in the shear-thinning fluids across the experimental range of shear rates $\dot \gamma$. (c) Average normalized swimming velocity of \textit{E.coli} perpendicular to the flow direction $V_y$ in the same fluids. Shear-thinning properties enhance the alignment of bacteria with the flow. Simulations match the experiments over the entire range of flow rates examined. Specifically, both simulations and experiments reveal a decrease in the lateral drift in the presence of the shear-thinning XG solution.}
 \label{Fig_3}
\end{figure}

\label{reorientationmodel}
The main goal of our model is to obtain expressions for the upstream ($V_x$) and lateral ($V_y$) swimming speeds in an applied flow. In this system, the upstream and lateral speeds, $V_x$ and $V_y$ respectively, are set by cell orientation with respect to the wall $\theta$ and the flow $\phi$ (see sketch in Fig.~\ref{Fig_3}a); see also \eqref{speed} in Methods. Our model therefore focuses on the mechanisms that set these angles and the corresponding reorientation rates $\Omega_{\theta,\phi}$ (see Eqs.~(\ref{oms} - \ref{reorevol}) in Methods). Here, we consider both hydrodynamic and steric interactions with the wall, as well as reorientation due to flow \cite{Kaya_2012} and shear-thinning viscosity effects \cite{Chen_2021}.

We model an elongated bacterium with a left-handed flagellum that swims forward at a constant speed $V_0$ along its axis. As a first approximation, the bacterium swims at a fixed height above the wall, $h$, independent of both fluid type and flow rate~\cite{Cao_2022}. We assume that, near the wall where a linear shear flow profile develops, the bacterium is advected at a speed $\dot \gamma h$. Thus, both quantities $V_0$ and $h$ can be extracted from the experimental data (see Methods) at zero and high shear rates respectively. We initially set $V_0$ constant at $10~\mu$m s$^{-1}$ and find a swimming height $h \sim 0.9 \mu m$ that is independent of fluid type. Note that while bacteria swimming speed ($V_0$) can increase in shear-thinning fluids, our results are relatively insensitive to the value of $V_0$. Furthermore, the changes in $V_0$ do not account for the observed changes in lateral speed $V_y$. 

\begin{figure*}
 \centering
 \includegraphics[width=6.4in]{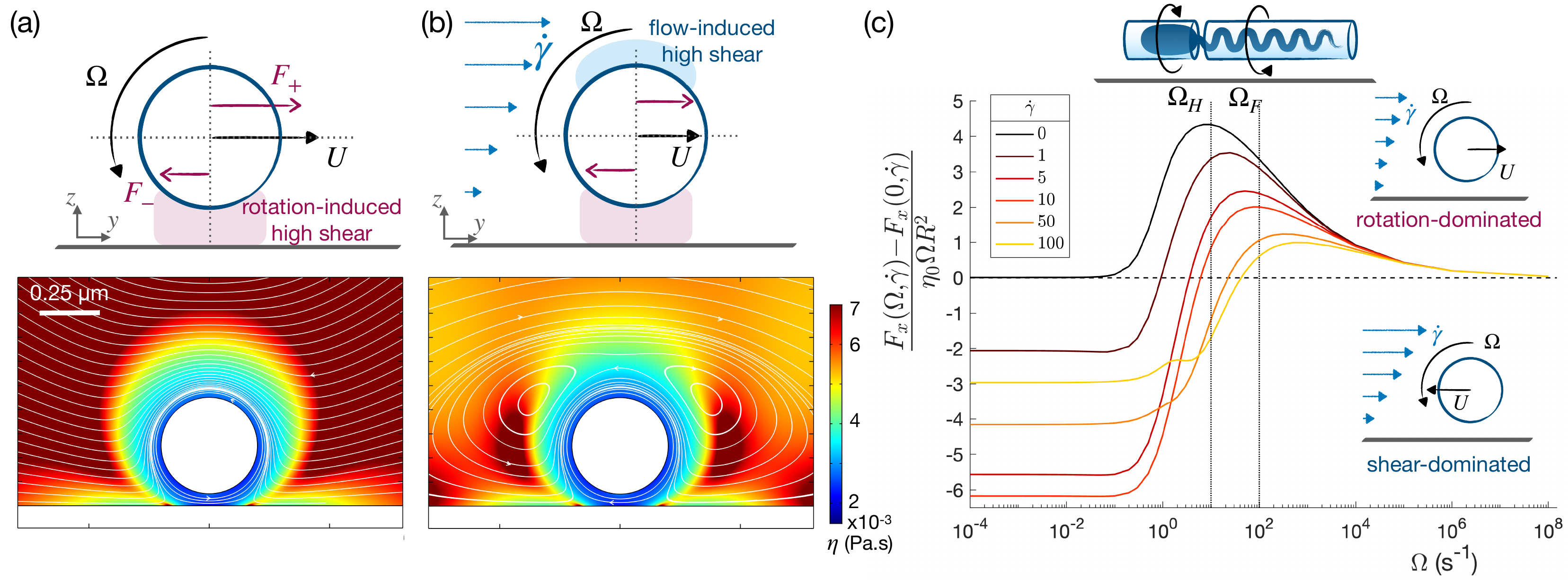}
 \caption{ The coupling between the rotation and translation of a cylinder in a shear-thinning fluid above a wall explains the reorientation of a bacterium with counter-rotating head and flagellum. 
 (a) Sketch (top) and simulated local viscosity (bottom) of a cylinder rotating at a rate $\Omega$ near a wall in a shear-thinning fluid. Confinement results in higher shear rates and, hence, lower viscosity $\eta$ on the lower side of the cylinder. As a result, the horizontal component of the force resisting the rotation $F_-$ is lower on the lower half than on the upper half $F_+$, consequently the rotation induces a translation to the right $U$. (b) When adding an external flow, high shears are instead induced on the upper half. The direction of the rotation-induced translation of the cylinder is therefore determined by a competition between these two effects. 
 (c) Difference between the scaled forces on the cylinder with ($F(\Omega,\dot\gamma)$) and without rotation  ($F(0,\dot\gamma)$) at different shear rates. When the shear rate dominates (low $\Omega$), the coupling occurs in the negative direction because the viscosity is overall lower above the cylinder. When rotation dominates (high $\Omega$), the viscosity is lower below, leading to a rotation-induced translation in the positive $x$-direction. The vertical lines indicate the rotation rates of the bacterium head $\Omega_\textsc{h}$ and flagellum $\Omega_\textsc{f}$ (sketched). }
 \label{Fig_4}
\end{figure*}

\subsection{No-flow Condition} 

\subsubsection{Newtonian fluids.} 
At a wall, bacteria preferentially swim parallel to the surface with a small tilt angle $\theta_0 \approx 10^\circ$ \cite{Kaya_2012,bianchi2017holographic}. This behavior is described in \eqref{oms} in Methods, which depends on $\theta$ and a prefactor that corresponds to the alignment strength (obtained from previous work \cite{Mathijssen_2019}). 

The motion of bacteria such as \textit{E.coli} near walls in Newtonian fluids is clockwise circular \cite{Lauga_2006}. This stems from the hydrodynamic interaction of the rotating head and the counter-rotating cylinder with the wall~\cite{Lauga_2006} (Fig.~\ref{Fig_4}c): in a viscous fluid, a 3D sphere rotating near a no-slip wall translates in the rolling direction. As the translations for the flagellum and the head go in opposite directions, the bacterium is reoriented and turns continuously to the right. We approximate this effect as a constant rotation rate due to chirality $\Omega_\phi^\textsc{c}$ (\eqref{omc} in Methods).  We note that, under flow, bacteria tend to swim upstream, but their chirality induces a drift to the right in the $+ y$-direction \cite{Hill_2007}, and their lateral velocity $V_y$ in a Newtonian fluid is positive (Fig.~\ref{Fig_3}c). 

\subsubsection{Shear-thinning fluids.} 
For a bacterium near a wall in a shear-thinning fluid, we don't expect the tilt angle to be modified~\cite{Cao_2022}. On the other hand, we do anticipate the curvature of the bacteria trajectories near walls to be modified by local viscosity gradients. Indeed, it has been previously shown that the translation of a cylinder rotating above a wall is modified by shear-thinning viscosity effects \cite{Chen_2021}.  Here, we numerically investigate the effect of shear thinning viscosity on the orientation of the bacterium in a quiescent fluid, following Ref.~\cite{Chen_2021}. For simplicity, we idealize the rotating bacterium as a rotating cylinder, as shown at the top of Fig.~\ref{Fig_4}c.

To minimize computation costs, we use finite element numerical simulations (see SI) to compute the respective forces, $|F_+|$ and $|F_-|$, on the upper and lower half of a cylinder rotating above a no-slip wall (Fig. \ref{Fig_4} and SI). In 2D, an infinite cylinder rotating in a quiescent Newtonian fluid exhibits zero translation close to a wall \cite{jeffrey_1981}; thus, our results account only for the shear-thinning effect which adds up to the previously described Newtonian chiral reorientation.  

In a quiescent shear-thinning fluid, strong velocity gradients develop between the wall and the cylinder, creating a relatively low viscosity region. Consequently, the bottom half of the cylinder experiences a lower drag force, $F_-$, compared to $F_+$ on the upper half; that is $|F_-|<|F_+|$.  Since the cylinder is force-free, the difference in forces leads to a net lateral translation $U$ in the direction opposed to rolling, which depends non-monotonically on the rotation rate $\Omega$~\cite{Chen_2021} (Fig.~\ref{Fig_4}, black line). 

Our simulations indicate that we should include in our model (\eqref{omst} in Methods) a CCW shear-thinning reorientation $\Omega_\phi^\textsc{st}$ that competes with the CW rolling effect $\Omega_\phi^\textsc{c}$ (Fig. \ref{Fig_3}a). Next, we investigate the case in which a flow is applied. 

\subsection{Applied Flow Condition}
\subsubsection{Newtonian fluids.} The driving effect of surface rheotaxis is the passive upstream reorientation of bacteria by the flow \cite{Kaya_2012, Kantsler_2014,brosseau2019relating}; this is the so-called weathervane effect. Because of the relatively high viscous drag and low flow speed near the wall, the bacterium head acts as an anchor, while its flagellum is allowed to reorient freely. The flagellum is advected by the flow like a weathervane and tends to align with it: this reorients the bacterium upstream and enables rheotaxis close to walls. The strength of the upstream alignment is fitted on the Newtonian data (\eqref{omwv}).

\subsubsection{Shear-thinning fluids} How is bacterial rheotaxis modified by shear-thinning viscosity effects? Our finite element simulations on a toy model representing a rotating cylinder advected in a flow hint at very weak effects of the shear-thinning fluid on the weathervane effect, which we consider independent of the fluid type (see SI). 

We now extend our finite element simulations for a rotating cylinder in a shear-thinning fluid to include an external shear flow $ + \dot \gamma$ or $ -\dot \gamma$ along the $y$-direction. The cylinder then acts as an obstacle that pushes streamlines upwards and leads to larger shear rates in the region above it (Fig.~\ref{Fig_4}b). It is now surrounded by two competing high-shear regions with low viscosity, one close to the wall controlled by the rotation and one above the cylinder set by the external flow. We identify the dominant effect (flow or cylinder rotation) by computing the net force acting on the cylinder ($F_{cyl}$) as a function of cylinder rotation rate ($\Omega$) for different values of $ + \dot \gamma$ in Fig.~\ref{Fig_4}c. For parameters values relevant to \textit{E. coli} rheotaxis, namely the rotation rates of the flagellum ($\Omega_\textsc{f} = 100\,\textrm{s}^{-1}$) and body ($\Omega_\textsc{h} = 10\,\textrm{s}^{-1}$) at intermediate shear rates ($\dot \gamma < 100\,\textrm{s}^{-1}$), the translation occurs in the positive $y$ direction, opposite to rolling. 

In our minimal model (Eq. 6 in Methods), we consider a linear decrease of the shear-thinning-induced reorientation $\Omega_\phi^\textsc{st}$ with the flow. Without flow, the shear-thinning reorientation compensates to the chiral one ($\Omega_\phi^\textsc{st} = -\Omega_\phi^\textsc{c}$), in agreement with experiments. This effect then decreases linearly with $\dot \gamma$, providing the flow-sensitivity fitting parameter for the shear-thinning case in Fig.~\ref{Fig_3}b. 

We neglect other possible but weaker contributions to the orientation \cite{Mathijssen_2019}, and only include rotational noise. Note that the only fitting parameters in our model are the strength of the weathervane effect for the Newtonian case and the shear-thinning reorientation and flow sensitivity (see \eqref{reorevol}). 

\subsection{Comparison between simulations and experiments}

\begin{figure}[t]
 \centering
 \includegraphics[width=\columnwidth]{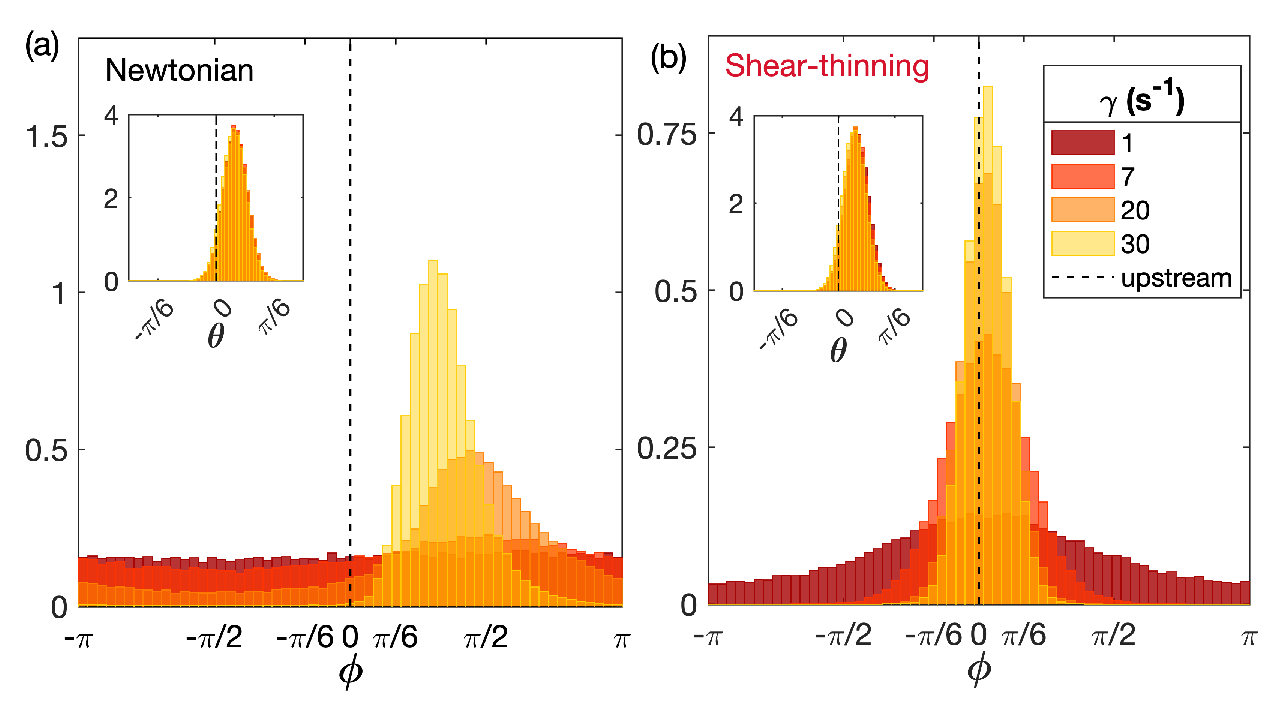}
 \caption{ Distribution of the lateral $\phi$ angle in our simulations for the Newtonian case (a) compared to the shear-thinning one (b) for different shear rates $\dot \gamma$. Insets show the tilt angle $\theta$, which depends weakly on the shear rate and fluid rheology. 
 At low shear rates ($\dot \gamma = 1\, \textrm{s}^{-1}$), the shear-thinning swimmers respond to the flow while the Newtonian ones are uniformly distributed. 
 At intermediate shear rates, the lateral distribution of shear-thinning swimmers shows a sharp peak close to the upstream position ($\phi = 0$), hence allowing for efficient rheotaxis. The alignment of Newtonian swimmers occurs instead at high shear rates ($\dot \gamma = 20, \, 30 \, \textrm{s}^{-1}$ ) when the external flow is already too strong to allow for upstream swimming.    }
 \label{Fig_5}
\end{figure}

Using our numerical simulations, we can now obtain time evolution and long-term averages for the swimming speed ($V_x$/$V_0$) and orientation ($V_y$/$V_0$) of our model bacteria. Figure~\ref{Fig_3}(b, c) shows a comparison between the experimental (symbols) and numerical (lines) results of $V_x$/$V_0$ and $V_y$/$V_0$, respectively, as a function of applied shear-rates. We start by noting that, for Newtonian fluids, both the model and experiments show minimal positive rheotactic behavior (Fig.~\ref{Fig_3}b, black line) across all shear rates. For shear-thinning fluids, the model is able to capture the deviation from the Newtonian case and the nonlinear rheotactic behavior as a function of shear rate (Fig.~\ref{Fig_3}b, red line); the largest bacterial upstream swimming speed can be as large as $40\%$ of its intrinsic swimming speed. We find that, in both experiments and simulations, the cross-over from upstream to downstream swimming occurs at shear rates that are an order of magnitude larger than in the Newtonian case. Similarly, our numerical results recover the bacterial swimming alignment (or lateral speed, $V_y$/$V_0$) behavior relatively well for both Newtonian (black line) and shear-thinning (red line) cases (Fig.~\ref{Fig_3}c); the simulations also capture the enhanced bacteria alignment with the applied flow in shear-thinning fluids (Figs.~\ref{Fig_1}d and~\ref{Fig_5}).

Overall, our numerical model captures the main features of the bacterial rheotatic behavior in Newtonian and shear-thinning fluids, including the strong bacteria rheotactic response at low and intermediate shear rates ($\dot\gamma$) and the reduction in lateral drift $\langle V_y \rangle $ for $\dot \gamma > 3 \, \textrm{s}^{-1}$. We also find, from our model, that the lateral angle ($\phi$) distributions (Fig.~\ref{Fig_5}) are significantly different in a shear-thinning fluid compared to a Newtonian one. On the other hand, the tilt angle ($\theta$) distributions remain nearly identical for the shear-thinning and Newtonian cases (Fig. \ref{Fig_5}, inset). 

These observations suggest that the dynamics are predominantly planar, or two-dimensional (2D), and allow us to calculate the critical shear rate, $\dot \gamma_{c,n}$, for positive rheotatic behavior, using a minimal model; for more details, please see SI (Section IIIB). This rheotactic transition can be described by an Adler equation~\cite{Adler_1946}. When the flow increases in a Newtonian fluid, the dynamics undergo a transition from loopy trajectories dominated by circular swimming to straighter (rheotactic) trajectories with a drift to the right reflecting the bacteria's chirality \cite{Goldstein_2009,tung2015emergence, Mathijssen_2019}. The critical shear rate, $\dot \gamma_{c,n}$, is reached when the weathervane effects compensate for the circular swimming (see methods and SI for more details); analysis shows that $\dot \gamma_{c,n} \approx 24$~s$^{-1}$ for Newtonian fluids. In shear-thinning fluids, the curvature of the trajectories is reduced and, as a result, the transition to rheotaxis occurs at a lower shear-rate value ($ \dot \gamma_{c,st} \approx 1 \, \textrm{s}^{-1}$). That is, the rheotaxis transition occurs at shear rates that are nearly an order of magnitude lower for shear-thinning fluids than for Newtonian fluids. Moreover, after this transition, the swimmers in the shear-thinning fluid align closely to the upstream direction, while in water (Newtonian), the rotation rate causes a strong drift to the right.  These combined effects enable rheotaxis at low and intermediate shear rates in a population of \textit{E. coli} in a shear-thinning environment. More broadly, we expect rheotaxis to be strongest for swimmers with straight trajectories or small curvature~\ref{Fig_4}b (see SI).

\section{Conclusion}

We find that shear-thinning viscosity significantly affects the rheotatic behavior of \textit{E. coli} near solid surfaces. Relative to Newtonian fluids, shear-thinning limits the chirality of the trajectories, which enables a bacteria rheotactic response to an imposed flow at relatively low shear rates; that is, the transition to bacterial rheotactic behavior happens at lower critical shear rates than in typical Newtonian fluids. After the transition, shear-thinning effects reduce the bacteria's lateral drift (to the right). Consequently, swimmers have sharper orientation distributions and are better aligned upstream, which ultimately increases their rheotactic ability in shear-thinning fluids. Conversely, a relatively large curvature of the trajectories in Newtonian solutions leads to a delay (in terms of $\dot\gamma$) in the transition to rheotaxis and to a drift to the right that weakens the upstream alignment. As a result, upstream swimming ($V_x < 0$) does not occur in the model and only marginally in experiments. Importantly, the fluid-dependent chiral rotation rate in a quiescent fluid is identified as a good predictor of swimmer population rheotactic ability, and we expect this finding to hold not only at the population but also at the individual level. 

Our results provide a better understanding of the behavior of swimming organisms in complex environments. Our work has important consequences for the study of medical, food, or wastewater treatment, and applications in designing medical devices that could prevent medical infections. In particular, we identify low curvature or, equivalently, high gyration radii, of bacteria trajectories near a surface in a quiescent fluid as a key predictive parameter of an efficient rheotaxis. As the curvature of the trajectories depends on the rheology of the surrounding fluid, we expect the rheotactic efficiency, and therefore the ability of a population to spread spatially and contaminate their environment, to greatly depend on the properties of the local fluid. Many polymeric solutions exhibit shear-thinning behavior, which we have shown here to enhance the rheotactic ability of \textit{E. coli}. The expectation is that other fluid rheological properties such as viscoelasticity, also common in many biological fluids, will also produce non-trivial rheotatic behavior (see for example, \cite{Cao2023}). More broadly, these biological complex fluids often possess a microstructure that responds nonlinearly to applied stresses, giving rise to their non-Newtonian flow behavior. Understanding the interplay between fluid microstructure and the flow fields collectively generated by swimming bacteria (near or away from walls) under flow is an important next step.

\vspace{60 pt}
\small

\section*{Methods}
\subsection*{Experimental Methods}

\paragraph*{Microfluidic device.}Experiments are performed in a microfluidic channel with dimensions of $900\,\mu$m width and $100\,\mu$m depth. The applied external volumetric flow, $Q$, is controlled through a syringe pump (Harvard Ph.D. 2000) over a range of flow rates $0\,\mu \textrm{L/hr} \leq Q \leq 100\,\mu $L/hr. A long delay line is incorporated at the start of the microfluidic device to provide high flow resistance and minimize the impact of pump-induced fluctuations and small bubbles as shown in Fig.~\ref{Fig_1}a. Our region of interest (ROI) is located on the bottom wall, away from the corners to avoid the effects of shear-trapping and edge currents \cite{Rusconi_2014, Figueroa_Morales_2015}. When flow is applied, the ROI is situated in the high-shear region of the fully developed flow.

\paragraph*{Polymeric fluids} The effects of fluid shear-thinning viscosity behavior on bacterial rheotaxis are investigated using two main polymeric fluids. Sear-thinning fluids with negligible levels of elasticity are produced using aqueous solutions of xanthan gum (XG), a semi-rigid polymer of molecular weight $M_w=2.0 \times 10^6$ and an overlap concentration $c^*=290\,$ppm. To minimize the effects of elasticity (see accompanying manuscript for more info on elastic effects), we perform experiments with dilute XG solutions ($c/c^* \leq 1$); four formulations are used: $c/c^*=0.2$, $c/c^*=0.4$, $c/c^*=0.6$, and $c/c^*=1.0$.  XG solutions show significant shear-thinning viscosity behavior (see SI, Fig.~S1).  We also experiment with aqueous solutions of carboxymethyl cellulose (CMC), a flexible polymer of molecular weight $M_w=7.0 \times 10^5$ and an overlap concentration $c^*=10^4\,$ppm. Elasticity is minimized by using a relatively low $M_w$ polymer in the dilute regime ($c/c^*=0.25$, $c/c^*=0.5$). Experiments with Newtonian fluids are performed with aqueous buffer solutions with shear viscosity ($\eta$) of 1 mPa s.  We viscosify the buffer solutions by adding Ficoll ($M_w=7.0 \times 10^4$, $R_g\approx 4.6\,$nm) at 10\% ($\eta \approx 3$ mPa s) and 15\% ($\eta \approx 5$ mPa s) in concentration by mass (see SI Fig.~S1). These Ficoll solutions serve as the control case.

\paragraph*{Rheology.} Fluids are characterized using a stress-controlled rheometer (TA instruments, DHR-3) with a $60\,$mm cone geometry (see SM). All rheological experiments are performed within the range of shear rates $\dot{\gamma}$ that the \textit{E. coli} experiences in the microfluidic channel. Viscosity data for the shear-thinning fluids (XG solutions) are fitted with the Carreau-Yasuda model \cite{Bird1987}

\begin{equation}
 	\eta\left(\dot{\gamma}\right)=\eta_{w}+\left( \eta_0-\eta_{w} \right) \left( 1+\lambda\dot{\gamma}\right)^{n-1}.
   	\label{CarYas}
\end{equation} 
This empirical model is known to capture the viscosity behavior of polymeric solutions for a wide range of shear rates. We find that the Carreau-Yasuda model provides an adequate description of our fluids for a range of $\dot{\gamma}=0.5-100\,$s$^{-1}$. Here, $\eta_0$ is the zero-shear ($\dot{\gamma}=0$) viscosity, $\eta_{w}=1\,$mPa$\,$s is the limit viscosity at high shear rates; here this value is set to the viscosity of the solvent, i.e., water. The time-scale $\lambda$ and the exponent $n$ characterize respectively the onset ($Cr=\lambda\dot{\gamma}>1$) and the magnitude of shear-thinning effects in the fluid. Note that XG solutions have values of  $\eta_0$ over an order of magnitude higher than water (or buffer solution). As expected, Newtonian Ficoll solutions display a constant viscosity in a dynamic setting (see SI Fig.~S1).

\paragraph*{Near-wall shear rate.} The flow shear rates ($\dot{\gamma}$) near the surface of the microfluidic channel are obtained using simple numerical simulations (\textsc{COMSOL} Multiphysics) for the various fluids and flow rates $Q$.  We obtain the values of $\dot{\gamma}$ for the XG solutions by using the Carreau-Yassuda fits to the rheology data in the simulations. That is, for each simulation, we input the flow rate and fluid properties obtained through fitting equation \ref{CarYas} to the rheological data. To extract the shear values from the 3D flow profile in the ROI, we calculate the average $\dot{\gamma}=\frac{\partial U_x}{\partial z}$ over a region of $z<7\,\mu$m.

\paragraph*{Cell tracking.} We track individual \textit{E. coli} \cite{patteson_running_2015} near the bottom wall of the microfluidic chamber in a 120\,$\mu$m$\times 120\,\mu$m field of view captured by a CCD camera (Sony XCD-SX90, 30 frames per second) and a microscope (Zeiss Z1, $20\times$ objective). We use \textit{E. coli} displacement data to calculate the cells' average swimming speed in both parallel $V_x$ and perpendicular $V_y$ directions to the flow. Bacteria that move less than one body length over the recording are excluded from the data analysis.

\subsection*{Model}

Surface rheotaxis stems from the interplay between
self-propulsion, the no-slip wall, and local flow. To under-
stand how shear-thinning enables upstream swimming,
we investigate how local differences in viscosity stemming
from shear-thinning affect these interactions. This section provides details on our model for the tilt and lateral orientation angles of the bacteria, $\theta$ and $\phi$, (Fig.~\ref{Fig_3}a). The vertical tilt angle is $\theta \in [-\pi/2; \pi/2]$, and the lateral angle with respect to the flow direction is $\phi \in [-\pi; \pi]$, so that $\phi = 0 $ represents upstream swimming, $\phi >0 $ denotes a drift to the right (i.e.,  $V_y >0$), and $\phi <0 $ denotes a drift to the left (i.e.,  $V_y <0$). 
The velocity of the swimming bacterium is then
\begin{equation}
\begin{split}
    V_x &=  \dot \gamma h - V_0 \cos \phi \cos \theta , \\
    V_y &= V_0 \sin \phi  \cos \theta .
\end{split}
\label{speed}
\end{equation}

The average swimming height ($h$) is extracted from experimental data using the swimmer's speed along the flow direction at large flow rates when $\dot \gamma h >> V_0 \cos \phi \cos \theta$.  We find that $h \approx 0.9~\mu$m, consistent with previous results \cite{Mathijssen_2019}.

Next, we provide minimal expressions for the mechanisms that set the tilt ($\theta$) and lateral angles ($\phi$) in a population of swimmers. Each torque that reorients a swimmer near a solid surface induces a rotation speed in $\theta$ and $\phi$, denoted respectively as $\Omega_{\theta}$ and $\Omega_{\phi}$. 
The combination of all the reorientation rates then sets the time evolution of the position of the swimmer as given by the ($\theta$,$\phi$) pair. 

\paragraph*{Wall alignment.} The preferred wall alignment rate, $\Omega^\textsc{s}_{\theta}$, is be described by the relationship \cite{Mathijssen_2019},
\begin{equation}
    \Omega^\textsc{s}_{\theta} = -w_s \sin\left(2 (\theta-\theta_0) \right),
    \label{oms}
\end{equation}
where $w_s = 4~{s^{-1}}$ is the magnitude of the reorientation rate and $\theta_0 = 10^\circ$ \cite{Mathijssen_2019}. 
In the absence of viscoelastic lift force, this preferred angle is expected to be independent of the fluid type.

\paragraph*{Circular swimming in Newtonian fluids.} We model the chiral interaction of the rotating flagellum and counter-rotating head with the wall as a constant rotation rate due to chirality,
\begin{equation}
    \Omega^\textsc{c}_{\phi} = w_\textsc{c} ,
    \label{omc}
\end{equation}
with $w_\textsc{c} = 1$ s$^{-1}$ for \textit{E. coli} ~\cite{Mathijssen_2019}. 

\paragraph*{Shear-thinning induced reorientation.} 
We estimate the translation induced by the shear-thinning behavior of the fluid is added as $\Omega^\textsc{st}$.
The swimmer is modeled using a rotating (infinite) cylinder for simplicity, which has been shown to provide useful insights into propulsion mechanisms \cite{Cao_2022}. 
Numerical simulations are performed using a shear-thinning Carreau fluid model fitted to our fluid rheological data to compute the resulting flow and force balance on the cylinder (see SI for details).
Cylinder rotation (above the wall) leads to a dimensionless net horizontal force $ F_{cyl} = F_y(\Omega) / \left(\eta_0 \Omega R^2 \right) $. Previous results \cite{Chen_2021} show a net cylinder translation in the direction opposite to rolling. Similar results are found here using the methods described in Ref. \cite{Chen_2021}, as shown in Fig.  \ref{Fig_4}c.

These results are extended to include flow effects, that is $\dot \gamma \neq 0$. We again consider a cylinder of radius $R$ held at a fixed position rotating parallel to a no-slip wall at a rate $\Omega$ and include an external shear rate $ + \dot \gamma$ or $ -\dot \gamma$ along the $y$ direction, as sketched in Fig.~\ref{Fig_4}b. Note that forces on a stationary cylinder (with no rotation) due to advection, $F(0,\dot \gamma)$, are already considered through the weathervane effect (see \eqref{omwv}). 

We compute the modified horizontal force on the cylinder $ F_{cyl} = \left(F_y(\Omega, \dot \gamma) - F_y(\Omega,0) \right) / \left(\eta_0 \Omega R^2 \right) $. We find a non-monotonic relationship between $F_{cyl}$ and $\Omega$, as well as two main regimes: shear- and rotation-dominated. In the shear dominate regime (relative low values of $\Omega$), resistance to rotation is lower on the upper half of the cylinder than in its lower half, $|F_+| < |F_-|$, and any rotation of the cylinder induces a rolling drift in the $-y$ direction. Importantly, for parameters values relevant to bacterial rheotaxis, namely the rotation rates of the bacterial flagellum ($\Omega_\textsc{f} = 100\,\textrm{s}^{-1}$) and head ($\Omega_\textsc{H} = 10\,\textrm{s}^{-1}$) and intermediate shear rates ($\dot \gamma < 100\,\textrm{s}^{-1}$), the translation occurs in the positive $y$ direction, opposite to rolling, as shown in Fig.~\ref{Fig_4}c. 

This new contribution is added to the model using a linear expression for the swimmer rotation rate such that: 
\begin{equation}
\Omega^\textsc{st}_{\phi} = - w_\textsc{st} \left(1 - \alpha \dot\gamma / \Omega_\textsc{f} \right) . 
\label{omst}
\end{equation}
The best fit to the experiments for the shear thinning effect is $w_\textsc{st} = w_\textsc{c} = 1 \,$s$^{-1}$, with a sensitivity to the flow $\alpha = 0.35$.

\paragraph*{Weathervane effect.} 
We approximate the swimmer reorientation due to the flow gradients at the wall by considering the movement of an object held at a fixed distance $L/2$ of an anchor at the wall and advected by the external linear shear flow (see SM) and add that this effect is only valid when the bacterium points to the wall, $\theta > 0$. We can describe the reorientation rates due to the weathervane effect, $\Omega^\textsc{wv}$, in the $\theta$ and $\phi$ directions as
\begin{equation}
\begin{split}
\Omega^\textsc{wv}_{\theta}& = - \dot\gamma w_\textsc{wv} (\sin \theta )^2 \cos \phi \, \mathbb{1}_{\theta > 0}  , \\  
\Omega^\textsc{wv}_{\phi} &= -\dot \gamma w_\textsc{wv} \sin\phi \tan \theta \, \mathbb{1}_{\theta > 0} .  
 \end{split}
 \label{omwv}
\end{equation}
The quantity $\Omega^\textsc{wv}$ scales linearly with $\dot{\gamma}$, as expected for Newtonian fluids. 

The effects of shear-thinning viscosity behavior on $\Omega^\textsc{wv}$ are not well known, so we investigate them by performing simulations of a two-dimensional (2D) cylinder being advected (near a wall) by a shear flow using the Carreau model in \textsc{Comsol}. The simulations do not show significant differences between the shear-thinning and Newtonian cases, particularly for the cylinder speed (see SM). 
Therefore, we set $w_\textsc{wv} = 0.24$ for all fluids investigated here; the $w_\textsc{wv}$ value is obtained by fitting the simulations from \eqref{reorevol} to the Newtonian experimental results.

\paragraph*{Noise} One could consider other contributions in our models such as a chiral interaction with the bulk flow and Jeffery orbits \cite{Kaya_2012, Mathijssen_2019}. Chiral interaction between the flagella and the bulk flow can, by themselves, lead to rheotaxis \cite{marcos2009, Marcos_2012}, but at the considered flow rate, they are small compared to wall-mediated ones. Jeffery orbits lead to a small oscillatory component in the trajectories \cite{Mathijssen_2019}, and they are not affected by shear thinning \cite{Abtahi_2019}. Thus, we neglect these two components.

Finally, we include rotational noise in our model through a Langevin equation adapted to our angular coordinates \cite{Raible_2004}, with strength $D_r = 0.057 \, \textrm{s}^{-1}$ \cite{Drescher_2011}. The flow-dependent reorientation rates of the bacterium are 
\begin{equation}
\label{reorevol}
    \begin{split}
        \dot \theta & =  \Omega^\textsc{s}_{\theta} + \Omega^\textsc{wv}_{\theta} (\dot \gamma) + D_r \tan \theta  +  \sqrt{2 D_r  } \, \text{d} \xi_\theta , \\
        \dot \phi & =  \Omega^\textsc{c}_{\phi} + \Omega^\textsc{wv}_{\phi} (\dot \gamma) + \Omega^\textsc{st}_{\phi} (\dot \gamma)  + \frac{\sqrt{2 D_r  } }{\cos \theta} \text{d}\xi_\phi. 
    \end{split}
\end{equation} 
where $\xi_\theta$ and $\xi_\phi$ are uncorrelated Gaussian white noises. 

\paragraph*{Model integration.} We use the above equations to obtain the dynamics (and values) of $\theta$ and $\phi$ for the experimental range of shear rates, $\dot \gamma$.  Similarly, we use \eqref{speed} (and distributions of $\theta$ and $\phi$) to obtain the average long-term velocities of the swimming bacteria as a function of $\dot \gamma$. 
Note that the only fitting parameters in our model are $w_\textsc{wv}$ for the Newtonian case and $w_\textsc{st}$ and $\alpha$ for the shear-thinning case. For the cases shown in Fig.~\ref{Fig_3}(b,c), $w_\textsc{wv} = 0.24 $ , $w_\textsc{st} = 1\,\textrm{s}^{-1}$, and $\alpha = 0.35$. The remaining parameters, namely $\theta_0 = 10 ^\circ$, $w_s = 4\,\textrm{s}^{-1}$ and $w_\textsc{c} = 1\, \textrm{s}^{-1}$, are obtained from the existing literature~\cite{Mathijssen_2019}. 

The quantitative features of our model, in particular, the predicted velocities ($V_x$, $V_y$) and the transition shear rate, are sensitive to the values of the reorientation parameters that we fit to the experiments, namely $w_\textsc{wv}$, $w_\textsc{st}$ and $\alpha$. Other model components are sensitive to the geometry of individual swimmers, such as the weathervane effect ($\Omega^\textsc{wv})$ that is expected to scale as $L^2$. Nevertheless, we find that model trends (Fig. \ref{Fig_3}b,c) are very robust to population diversity and changes in parameters. 
 
\paragraph*{Transition to rheotaxis} The critical Adler shear rate $\dot \gamma_{c,n}$ is reached when the chirality and weathervane rotations are similar such that $\dot \gamma_{c,n} \sim w_\textsc{c} / ( w_\textsc{wv} \sin \theta_0) $, with a deterministic drift to $\phi = \text{arcsin}\left( \dot \gamma_{c,n} / \dot \gamma \right)$ (see SI for the analytical analysis). In our model, $\dot \gamma_{c,n} \approx 24 \, \textrm{s}^{-1} $ but the noise in the system results in a progressive rather than sharp transition with a broad distribution of $\phi$ \cite{Goldstein_2009}. Increasing $\dot\gamma$ beyond the transition range and above experimental values ($\dot \gamma > 25\, \textrm{s}^{-1}$) allows the weathervane effect $\Omega^\textsc{wv}$ to overcome the constant chiral reorientation $\Omega^\textsc{c}$.  For high $\dot\gamma$, the distribution of $\phi$ narrows and tends to lower values, reflecting a strong upstream alignment of the swimmers. The flow at this point is too high to allow for a net upstream displacement. 

In a shear-thinning fluid, the chiral rotation above the surface is hindered by shear-thinning-induced reorientation. Since $|w_\textsc{c} - w_\textsc{st}| < D_r$ in \eqref{reorevol}, $\phi$ is dominated by the fluctuations and the quiescent trajectories are noisy. Similarly, at low $\dot\gamma$, noise dominates and the distribution of $\phi$ is broad despite a small upstream bias. As $\dot\gamma$ increases, the weathervane effect becomes prominent and encourages visible upstream alignment as early as $\dot{\gamma} = 1 \, \textrm{s}^{-1}$. Since the chirality is hindered by the shear-thinning reorientation, the Adler transition occurs for lower values of $\dot\gamma$ than in the Newtonian case, $\dot \gamma_{c,st} \sim (w_\textsc{c} - w_\textsc{st}) / ( w_\textsc{wv} \sin \theta_0)$, with $\dot \gamma_{c,st} \ll \dot \gamma_{c,n}$. With our parameters, $ \dot \gamma_{c,st} < 1 \, \textrm{s}^{-1}$, meaning that the transition is effectively controlled by the noise in the system in our simulations. Consequently, the $\phi$ distributions are much sharper and closer to direct upstream swimming ($\phi = 0$), although a weak bias to the right remains in Fig.~\ref{Fig_3}a (see SM). The swimmers can now effectively move upstream, as observed experimentally and predicted in our simulations~\ref{Fig_4}b. 

\normalsize	

\vspace*{12pt}
\begin{acknowledgments}
B.O.T.M. and P.E.A. acknowledge support from the National Science Foundation (NSF) Grant DMR-1709763.  A.T. received support from the Simons Foundation through the Math + X grant awarded to the University of Pennsylvania. A.J.T.M.M. acknowledges funding from the United States Department of Agriculture (USDA-NIFA AFRI grants 2020-67017-30776 and 2020-67015-32330), the Charles E. Kaufman Foundation (Early Investigator Research Award KA2022-129523) and the University of Pennsylvania (University Research Foundation Grant and Klein Family Social Justice Award).
\end{acknowledgments}

\bibliography{reference_list}

\setcounter{section}{0}
\setcounter{figure}{0}

\pagebreak 
\newpage 

\onecolumngrid

\vspace*{20mm}

\pagebreak

\vspace*{10mm}

\begin{center}
    {\Large{ \textbf{\textsc{Supplementary Information}} \\  \vspace{12pt} Enhancement of Bacterial Rheotaxis in Non-Newtonian Fluids} }
\end{center}

\vspace*{18 pt}

\section{Fluid rheology}
\begin{figure*}[h!]
 \centering
 \includegraphics[width=\columnwidth]{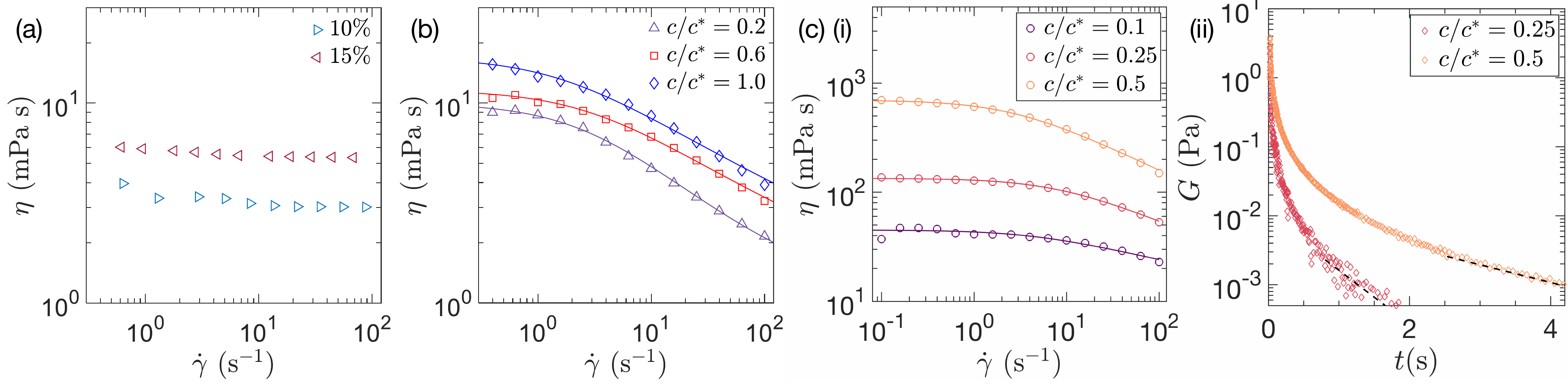} 
 \caption{(a) Experimental measurements of the steady shear viscosity $\eta$ as a function of shear rate $\dot{\gamma}$
 for $10\%$ and $15\%$ Ficoll solutions. (b) Experimental measurements of the steady shear viscosity $\eta$ of XG as a function of shear rate $\dot{\gamma}$ at $c/c^*=0.2$ (violet), $c/c^*=0.6$ (red), and $c/c^*=1.0$ (blue). (c) (i) Steady shear viscosity of CMC solutions and (ii) stress relaxation at $c/c^*=0.1$ (purple), $c/c^*=0.25$ (pink), and $c/c^*=0.5$ (orange).  
 The solid lines for $\eta$ are obtained from the best fits to the Carreau-Yasuda equation. 
 }
 \label{SI_Fig_1}
\end{figure*}

We measure the steady shear viscosity $\eta$ of Ficoll, xanthan gum (XG), and carboxymethyl cellulose (CMC) solutions using a cone-and-plate rheometer across a range of shear rates $\dot \gamma$ from $0.5$ to $100 \,$s$^{-1}$ (Fig.~\ref{SI_Fig_1}).  
In all fluids, the steady shear viscosity increased with the polymer concentration. The viscosities of the Ficoll and XG solutions are within the same order of magnitude, while that of CMC is higher.
Ficoll solutions exhibit a constant viscosity across this range of shear rates (Fig.~\ref{SI_Fig_1}a). On the other hand, both XG and CMC solutions displayed shear thinning behavior within the applied range of shear rates, as shown in SM Fig.~\ref{SI_Fig_1}b and c respectively. The CMC fluid also has some elasticity, as shown in Fig.~\ref{SI_Fig_1}c(ii). Using $G(t) = G_0e^{-t/\lambda}$, we extract the relaxation times $\lambda = 0.56s$ for $c = 0.25c^*$ and  $\lambda = 1.7s$ for $c = 0.5c^*$. 

\section{\textit {E. coli} rheotaxis in different fluids}
\subsection{Absence of flow}

\begin{figure*}[!ht]
 \centering
 \includegraphics[height=6cm]{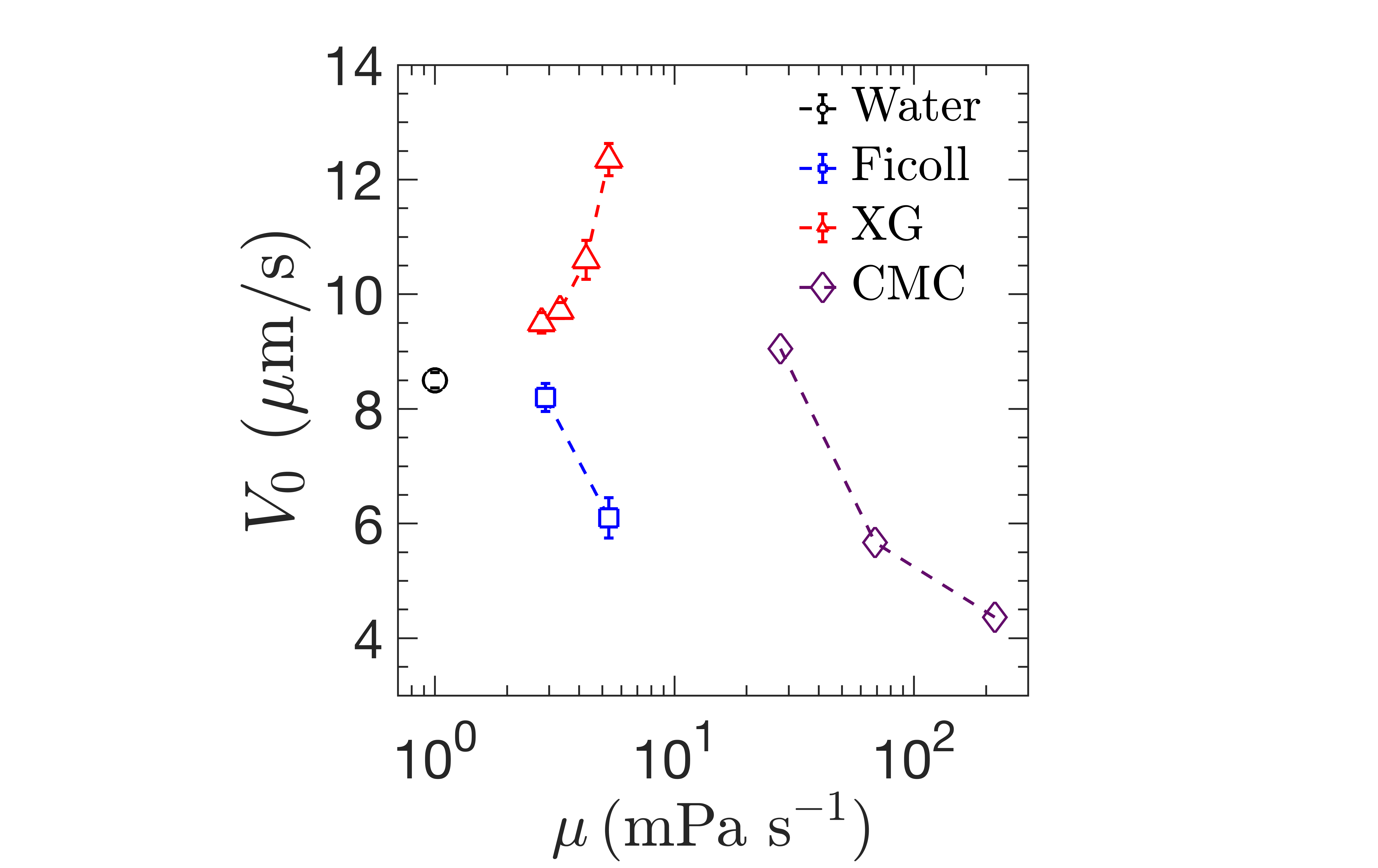}
 \caption{ Mean swimming speed $V_0$ of the bacterial population as a function of effective viscosity $\mu$. The effective viscosity $\mu$ was obtained as an average in the range of $\dot{\gamma}\approx 10-100 \,$s$^{-1}$, corresponding to the range of rotation rates of the head and flagella of \textit{E. coli}. Mean swimming speeds were obtained for four different fluids, water (black), Ficoll with concentrations of $10\%$ and $15\%$ (blue), XG solutions at $c/c^*=0.2$, $c/c^*=0.4$, $c/c^*=0.6$, $c/c^*=1.0$ (red) and CMC solutions at $c/c^*=0.1$, $c/c^*=0.25$, $c/c^*=0.5$ (purple). 
 }
 \label{SI_Fig_2}
\end{figure*}

We first compare the swimming velocities of \textit{E. coli} in the different fluids we used. 
To do so, we first compute the mean swimming speed $V_0$ of the bacterial population from the quiescent tracks. 
We then define an effective average viscosity by averaging the viscosity measurements (SM Fig.~\ref{SI_Fig_1}) between $\dot \gamma=10$ and $100\,$s$^{-1}$, which corresponds to the typical range of \textit{E. coli} flagellar rotation frequency \cite{Berg_2007,sowa_berry_2008}. This allows us to compare the swimming velocities for different fluids in Fig.~\ref{SI_Fig_2} and to isolate the effects of a shear-thinning viscosity on the swimming speed. 

The swimming velocity $V_0$ increases with dilute concentrations of XG in the shear-thinning fluids. In contrast, in Newtonian Ficoll, $V_0$ decreases monotonically with the Ficoll concentration, showing that, as expected, a Newtonian viscosity increase slows down the propulsion of the bacteria.

\subsection{Rheotaxis in flow}

\begin{figure*}[!ht]
 \centering
 \includegraphics[width=4.5in]{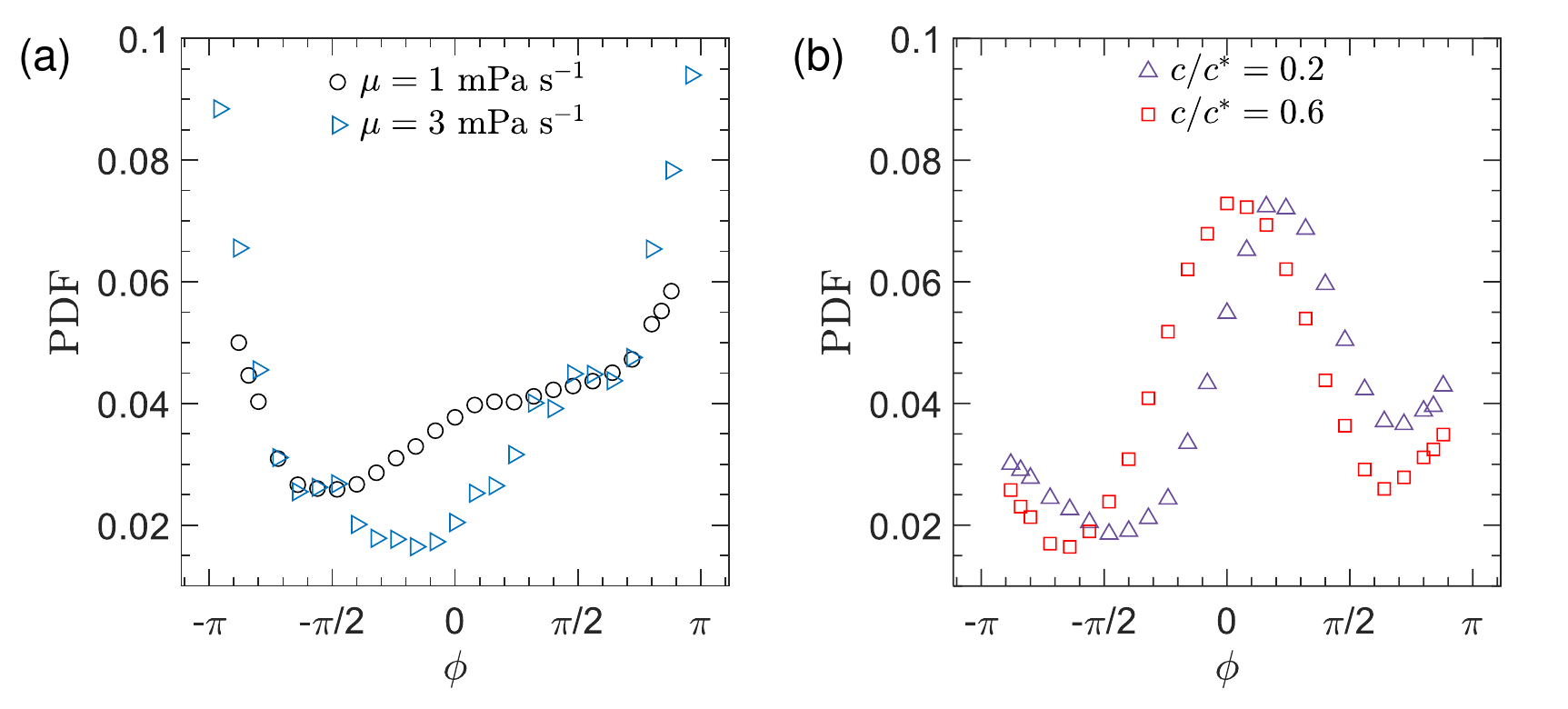} 
 \caption{ (a) PDFs depicting $\phi$ in Newtonian fluids: water (black) and a $10\%$ by weight solution of Ficoll (blue). (b) PDFs depicting $\phi$ in shear thinning solutions: XG at $c/c^*=0.2$ (violet), and $c/c^*=0.6$ (red).
 }
 \label{SI_Fig_3}
\end{figure*}

We analyzed the trajectory orientation of \textit{E. coli} bacteria relative to the flow direction, $\phi$, in Newtonian and shear-thinning fluids by generating probability density functions (PDFs) (SM Fig.~\ref{SI_Fig_3}). The data shows a slight clockwise bias in their swimming direction due to \textit{E. coli} chirality. As fluid viscosity is increased (using Newtonian Ficoll solutions), more bacteria is found to be advected downstream and fewer trajectories are found to go against the flow.  In shear-thinning fluids (XG solutions), however, we find more bacterial trajectories to be oriented against the flow main direction, despite having comparable or even higher viscosity than Newtonian fluids.

\subsection{Rheotaxis in a Newtonian fluid and in a shear-thinning, viscoelastic fluid}

\begin{figure*} [!ht]
 \centering
 \includegraphics[width=5in]{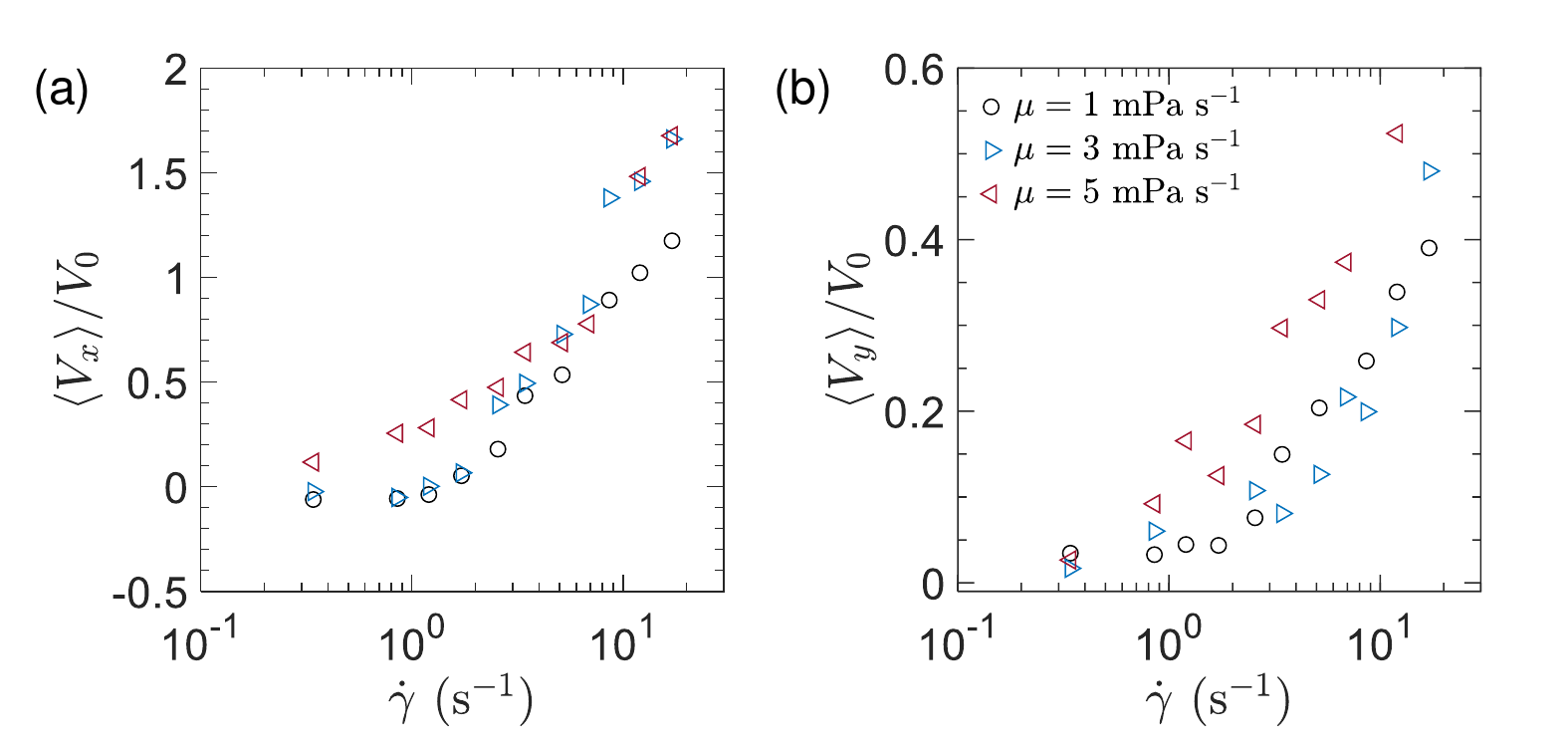}  
 \caption{(a) Experimental measurements of the average swimming velocity parallel to the flow direction, $V_x/V_0$, where $V_0$ is the average swimming speed of bacteria without an external flow ($Q=0\,\mu$L/hr) as a function of shear rate $\dot{\gamma}$. Results are shown in water (black) and Ficoll solutions with concentrations at $10\%$ (blue), and $15\%$ (red) by weight. (b) Average swimming velocity perpendicular to the flow direction, $V_y/V_0$, as a function of shear rate. }
 \label{SI_Fig_4}
\end{figure*}

We aim to investigate how different flow rates (and shear rates) affect the rheotatic behavior of bacteria near walls. Our approach involves computing the bacteria's average swimming speed parallel $V_x$ to the flow direction as a function of shear rates $\dot{\gamma}$, as shown in SM Fig.~\ref{SI_Fig_4}. In a Newtonian fluid, our experimental results reveal a limited amount of rheotaxis at low shear rates. As the shear rates are increased, the average bacterial velocities show a net displacement downstream. We conducted analogous experiments using a Newtonian medium with higher viscosity, revealing that for viscosities $\mu \geq 2\,$mPa$\,$s$^{-1}$, upstream swimming was weakly impeded rather than improved (Fig.~\ref{SI_Fig_4}a). The normalized lateral drift velocity $v_y/V_0$, on the other hand, varies only weakly with changes in Newtonian viscosity (Fig.~\ref{SI_Fig_4}b).

\begin{figure*} [!ht]
 \centering
 \includegraphics[width=5.5in]{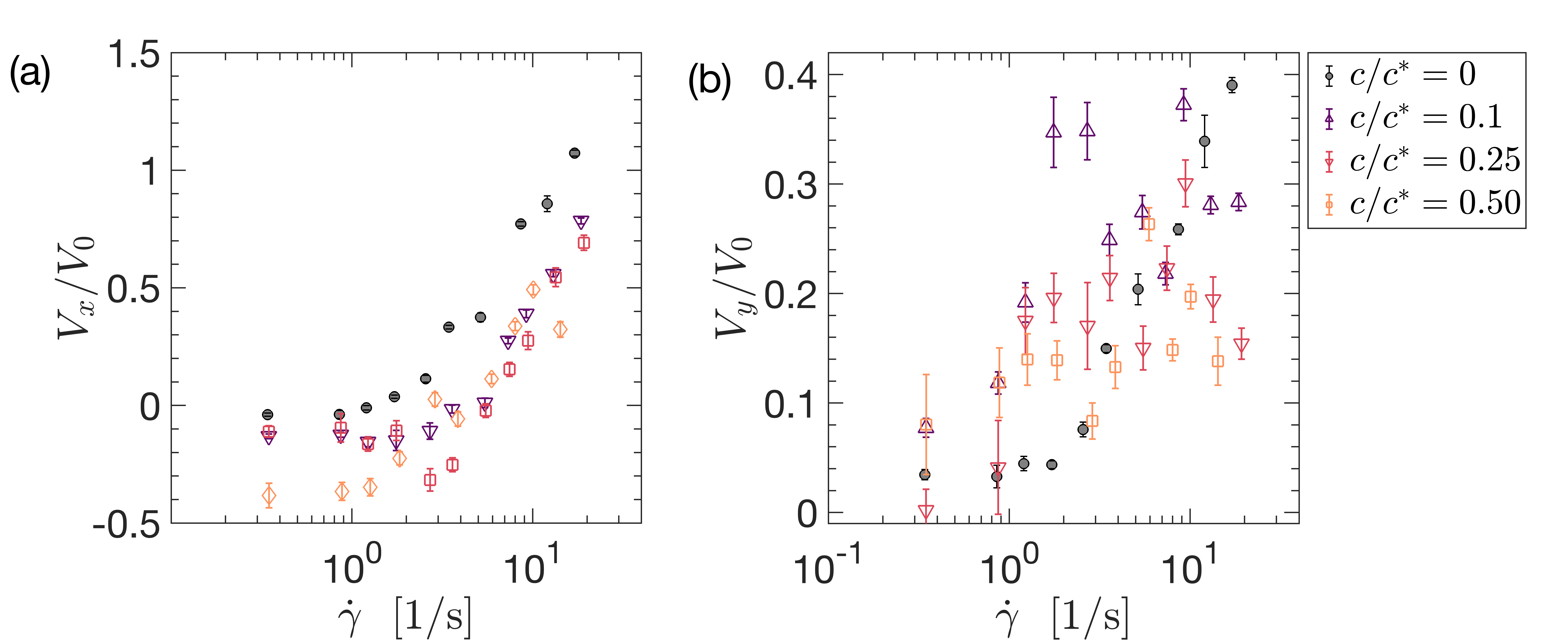} 
 \caption{(a) Experimental measurements of the average swimming velocity parallel to the flow direction, $V_x/V_0$, where $V_0$ is the average swimming speed of bacteria without an external flow ($Q=0\,\mu$L/hr) as a function of shear rate $\dot{\gamma}$ in water (black) and different concentrations of CMC (empty symbols). }
 \label{SI_Fig_CMC}
\end{figure*}

We also show that the enhanced rheotaxis is a generic effect rather than a xanthan gum-specific one, we also carry out the rheotaxis experiments in carboxymethyl cellulose (CMC), which produces shear-thinning, weakly viscoelastic fluids (see Fig.~\ref{SI_Fig_1}c).  
The normalized rheotactic speed against the flow $V_x/V_0$ is again strongly enhanced in CMC, suggesting that rheotaxis is generically enhanced in non-Newtonian fluid (Fig.~\ref{SI_Fig_CMC}a). On the other hand, the trend in lateral velocity is different than in the shear-thinning and non-elastic suspensions. This hints at an additional effect of elasticity in the flow response of \textit{E. coli}, which is outside the scope of this paper~\cite{Cao2023}.

\section{Model for swimmer reorientation in a shear flow}
\subsection{Weathervane effect}

\begin{figure*}[!ht]
\centering
\includegraphics[width=.6\textwidth]{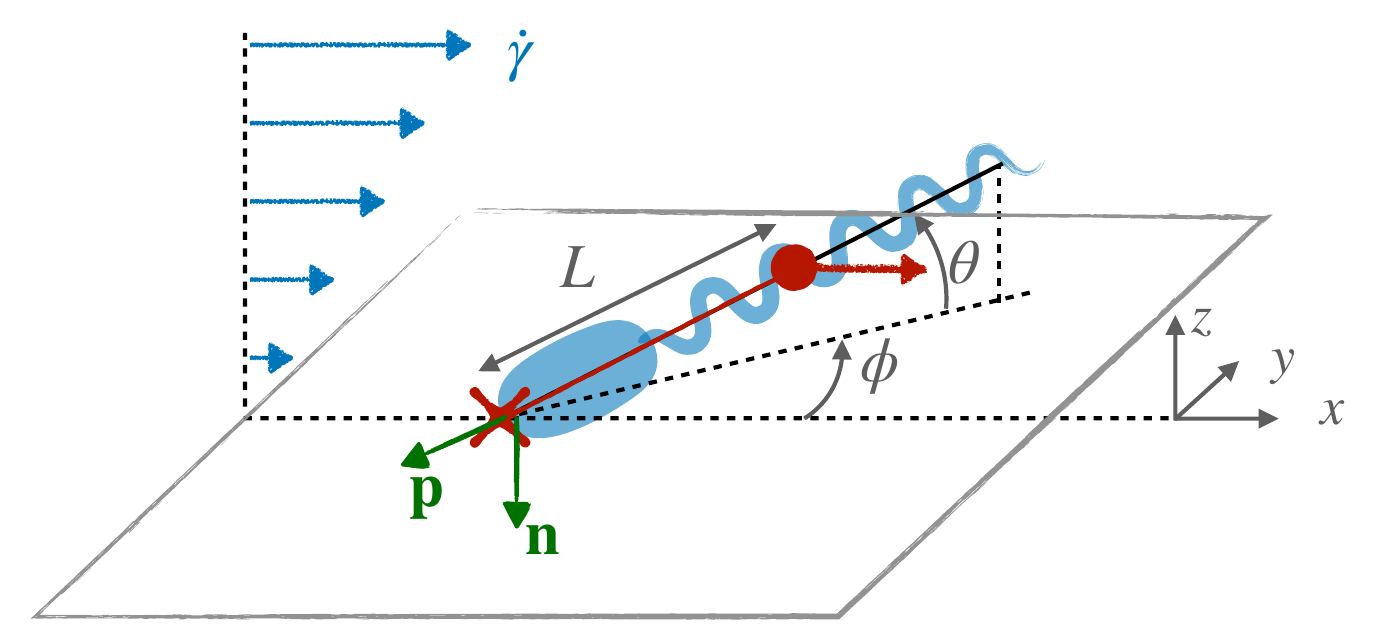}
\vspace{-2mm}
\caption{ Sketch of our model for the weathervane effect: a particle is held at a fixed distance $L$ from an anchor at the wall and advected by an external shear flow of strength $\dot \gamma$.}
\label{fig:SIwvsketch}
\end{figure*}

To understand how the weathervane effects depend on a swimmer's orientation, we consider a sphere held at a constant distance $L$ from an anchor at the wall and exposed to an external shear flow. The flow on the sphere is $ ( - L \bm{p} \cdot \bm{n})$, where $\bm{p}$ is the orientation of the swimmer and $\bm{n}$ the normal to the wall [see SM Fig.~\ref{fig:SIwvsketch}]. The moment of the force from the flow at the sphere is then
\begin{equation}
    \bm{m} = \dot \gamma (L \bm{p} \cdot \bm{n} ) L (\hat{\bm{x}} \times \bm p)  \quad \textrm{where} \quad 
    \bm p =   \begin{pmatrix} - \cos \theta \cos \phi \\ -  \cos \theta \sin \phi \\ - \sin \theta \end{pmatrix} . 
\end{equation}
The rotation rate of the sphere scales as 
\begin{equation}
       \bm \omega = \tilde \omega \dot \gamma L^2  \sin \theta  \begin{pmatrix} 0 \\  \sin \theta  \\ - \cos \theta \sin \phi  \end{pmatrix}, 
\end{equation}
For a small time step $\dd t$, we evaluate the changes $\dd \theta$ and $\dd \phi$ in tilt and lateral angles. We use that $\bm p(t+ \dd t) = \bm p(t) + \dd t \, \bm \omega \times \bm{p}(t)$, and develop the expressions to the first order in $\dd \theta$ and $\dd \phi$ as
\begin{equation}
    \bm p (t+ \dd t) - \bm p(t) = \begin{pmatrix}
        - \cos (\theta + \dd \theta)  \cos (\phi + \dd \phi) +  \cos \theta \cos \phi \\  - \cos (\theta + \dd \theta)  \sin (\phi + \dd \phi) \cos \theta \sin \phi \\ - \sin(\theta + \dd \theta) + \sin \theta
    \end{pmatrix}
    = \begin{pmatrix}
       \dd \theta  \sin \theta \cos \phi  + \dd \phi  \cos \theta \sin \phi \\  \dd \theta \sin \theta \sin \phi  - \dd \phi \cos \theta \cos \phi  \\ - \dd \theta \cos \theta 
    \end{pmatrix}
\end{equation}
and 
\begin{equation}
    \dd t \, \bm \omega \times \bm p =  \dd t\, \dot\gamma w_\textsc{wv} \sin \theta \begin{pmatrix}
        - 1 + ( \cos \theta \cos \phi)^2 \\ (\cos \theta)^2 \sin \phi \cos \phi \\  \sin \theta \cos \theta \cos \phi
    \end{pmatrix} .
\end{equation}
This provides three equations for $\dd \theta$ and $\dd \phi$, which we rearrange to get 
\begin{equation}
     { \begin{aligned}
     \dd \theta & = -  \dd t \, \dot\gamma w_\textsc{wv}  (\sin \theta)^2 \cos \phi ,  \\ 
     \dd \phi & = - \dd t \, \dot\gamma w_\textsc{wv}  \tan \theta \sin \phi .
\end{aligned} } 
\end{equation}
The linear coefficient $ w_\textsc{wv}$ is the fitting parameter of our Newtonian simulations.

\subsection{The radius of gyration controls the transition from loopy to rheotactic trajectories}

\begin{figure}[h!]
    \centering
    \includegraphics[width = \columnwidth]{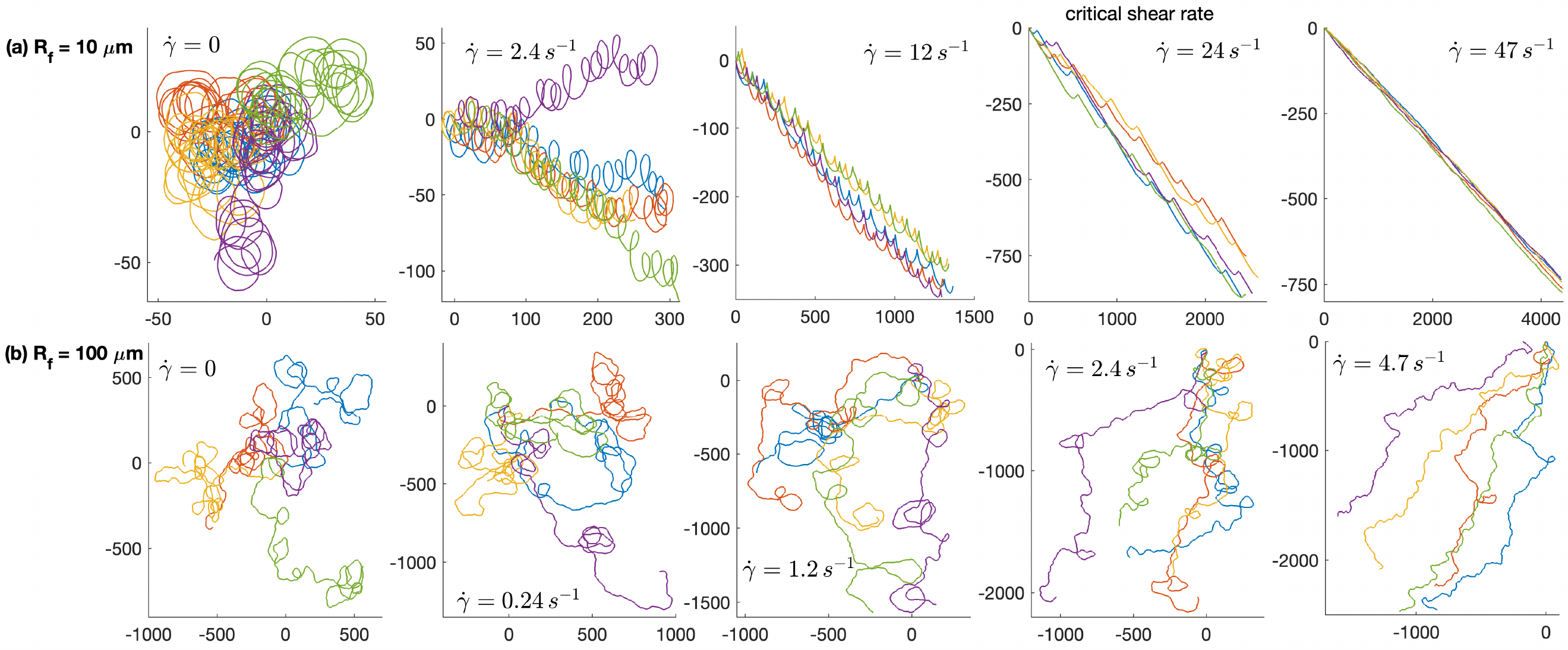}
    \caption{ Transition to rheotactic behavior for different gyration radius. (a) Trajectories for $125\, s$ or the parameters fitted to the experiments in water, the transition occurs at high shear rates with a large drift CW. (b) Trajectories for $500 \, s$ and larger gyration radii, as in the shear-thinning case. The transition to rheotaxis occurs for lower shear rates, and upstream swimming is robust. The role of noise on setting the trajectories around the transition is more pronounced.  }
    \label{adlertraj}
\end{figure}

Our simulations suggest that the chirality of a bacterium's trajectory above the wall is a key parameter to evaluate the rheotactic ability of an individual or a population. Without flow, this chirality can be evaluated with the gyration radius $R_f$. In particular, a higher gyration radius, or equivalently a reduced curvature, as in a shear-thinning fluid, appears to induce an earlier transition to a rheotactic behavior. This is evident in the sample trajectories for two different radii shown in Fig.~\ref{adlertraj}.

To understand how the rheotactic transition is set, let us consider a simplified two-dimensional model for the swimmer trajectories. Here, we ignore variations in their tilt angle, which we set at $\theta = \bar \theta$, and focus only on the lateral orientation $\phi$. 
The radius of gyration $R_f$ depends on the individual bacterium and the external fluid properties, and with parameter in our model writes
\begin{equation}
    R_f = v_0 \left(\omega_c - \omega_{st}(0) \right)^{-1}. 
\end{equation}
In particular, our experiments and analysis corroborate previous analytical results \cite{Cao_2022} that show that the quantity $R_f$ increases sharply in shear-thinning fluids. Note that, for simplicity, we neglect variations of $\omega_{st}$ with flow shear rates, that is $\alpha=0$.

The weathervane effect strength can be rescaled as $\tilde w_{vw} =  w_{vw} \tan \bar \theta$, and the bacteria orientation $\phi$ now reduces to
\begin{equation}
    \frac{ \partial {\phi}}{\partial t} = \frac{v_0}{R_f} - \dot \gamma \tilde \omega_{wv} \sin \phi + \sqrt{2 D_r} \xi_\phi . 
\end{equation}
This is a stochastic Adler-type equation, which denotes the ability of coupled oscillators to synchronize \cite{Adler_1946, Niedermayer_2008, Goldstein_2009}. In the context of rheotaxis, the transition to synchrony stands for the ability of the swimmers to go from loopy to oriented rheotactic trajectories. 

In the deterministic case ($D_r = 0)$, this transition occurs at a critical shear rate $\dot \gamma_c$ 
\begin{equation}
    \dot \gamma_c = \frac{v_0}{R_f \tan \bar \theta w_{wv}} 
\end{equation}
Below $\dot \gamma_c$, the swimmers perform loops, which hinders their ability to respond to external flows. The above expression indicates that the critical transition shear rate, $  \dot \gamma_c $, is inversely proportional to $R_f$. Thus, bacteria with large values of $R_f$ and correspondingly straighter trajectories in quiescent fluids can transition to rheotaxis at lower shear rates.

\begin{figure}[b!]
    \centering
    \includegraphics[width = 135mm]{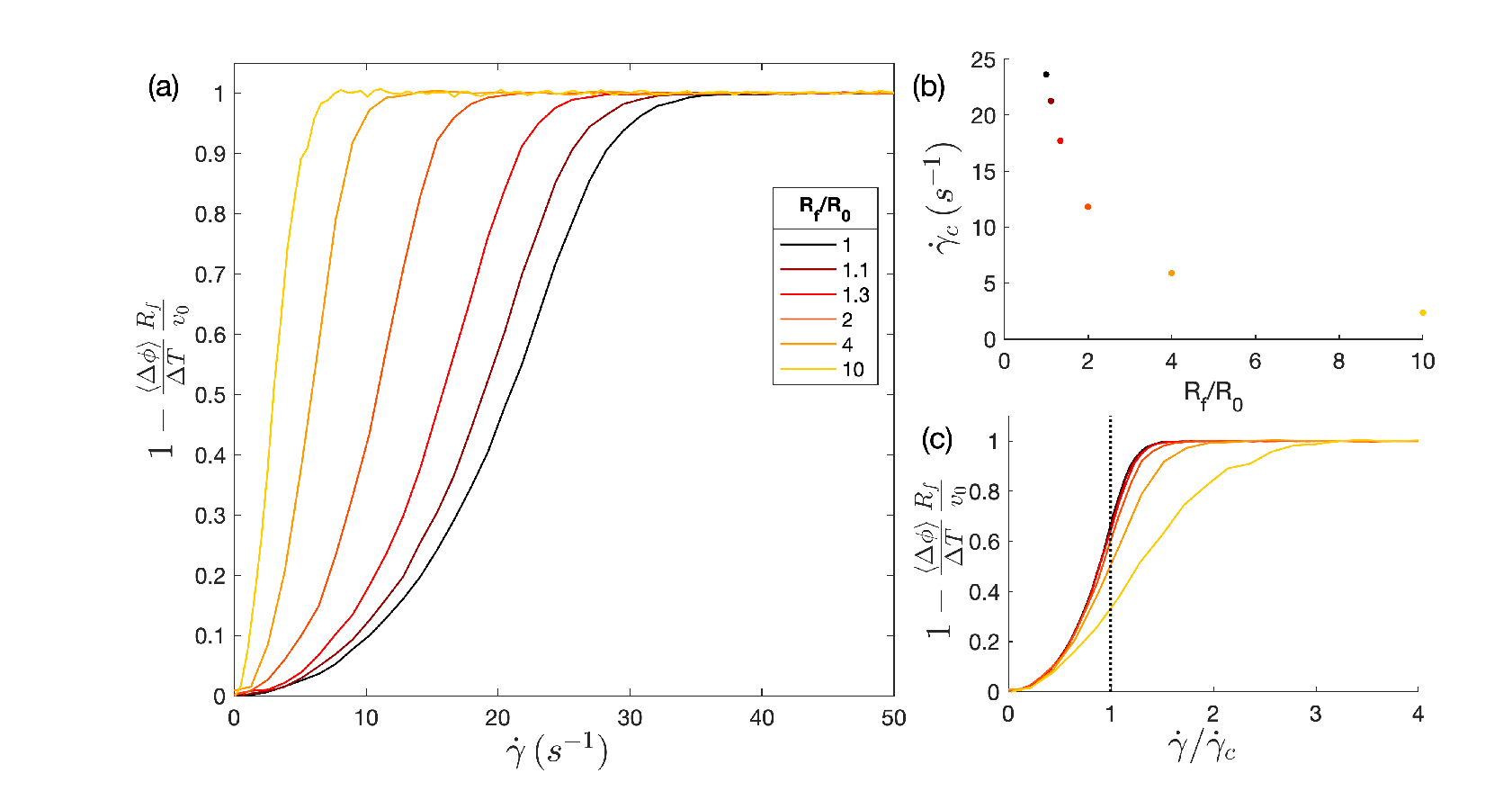}
    \caption{ (a) Transition to rheotactic behavior for different gyration radii $R_f$. The scaled reorientation rate $ 1 - R_f \langle \Delta \phi \rangle/ \Delta T /v_0 $ is $0$ for a circular swimmer and $1$ for a straight one.  (b) Critical shear rate $\dot \gamma_c$ as a function of the gyration radius $R_f$.  (c) Rescaling the reorientation rate shows that the transition to straight swimming occurs at $\dot \gamma_c$. The delayed transition for high $R_f$ comes from the noise in the system.}
    \label{adlertransition}
\end{figure}

To quantify this transition from circular to straight swimming in our simulations, we focus on the reorientation rate of the swimmers in the steady state, $\langle \Delta \phi \rangle/ \Delta T  = \langle (\phi(\Delta T) - \phi(0)) \rangle/ \Delta T  $, where $\Delta T$ is a time interval much larger than the typical reorientation timescales $ \Delta T \gg w_{c}^{-1},w_{wv}^{-1}$. 
For a circular swimmer, the reorientation rate and equal to $\Delta \phi/\Delta T = \omega_c = v0/R_f$, while for a straight one, $\Delta \phi/\Delta T=0$. 
Fig.~\ref{adlertransition}a shows the rescaled reorientation rate of the swimmers with time $ 1 - R_f \langle \Delta \phi \rangle/ \Delta T /v_0 $, so that $0$ corresponds to a circular swimmer and $1$ to a straight swimmer.  
The transition from circular to straight swimming occurs at $ \dot \gamma_c$ (Fig.~\ref{adlertransition}c), even in simulations where $\theta$ can vary. 
This shows that our simple 2D Adler model provides a good description of our 3D simulations.

The ability to respond to low flow rate for bacteria moving in shear-thinning fluid, and thus having straighter trajectories, is a strong advantage for rheotaxis: it allows the swimmers to swim against the flow before the flow is strong enough to advect them. 

Beyond the transition, the trajectories are still oscillatory but do not exhibit loops: the swimmers respond to flow and are oriented upstream around an angle $\phi^*$ given by
\begin{equation}
    \phi^* (\gamma) = \asin \left[\frac{\dot \gamma_c}{\dot \gamma} \right]. 
\end{equation}
The swimmers first exhibit a strong bias CW, before progressively aligning against the flow at high shear rates. 
Fig.~\ref{adlertraj}(a) shows typical trajectories of swimmers with increasing flow rates.

In the stochastic case, noise also affects the trajectories as well as the transition. The relative importance of reorientation because of noise is stronger for relatively higher gyration radii, and it then delays the transition as shown in Fig.~\ref{adlertransition} for $R_f = 10$.

\subsection{Simulations for a rotating cylinder above a wall}
To understand the behavior of a rotating cylinder in an external flow close to a wall, we perform simulations of the Stokes equations in COMSOL Multiphysics. 
We use a 2D rectangular domain of size $L\times H = 500 \times 1500 R$. 
We prescribe a no-slip boundary condition on the bottom wall, while the upper wall is translating at $ \dot \gamma H $ to impose the shear flow. 
We also impose the rotation speed $\Omega$ of the cylinder and extract the force imposed on the cylinder in the horizontal direction, $F_x(\Omega,\dot \gamma)$. We then subtract the horizontal drift stemming from advection at $\Omega = 0 $, $F_x(0 ,\dot \gamma)$, to evaluate the effect of the translation-induced drift. 
We refine our free-triangular mesh around the cylinder, and our system finally has $4 \times 10^5$ degrees of freedom for our default parameters. 
We validate our results without an external shear-flow against the work of \citet{Chen_2021}. 

Unless stated otherwise, we use $R = 0.2 \si{\mu m}$ and $h = 1.25 R $ for the geometry. 
For the shear thinning fluid, we use the Carreau model from \textsc{Comsol}, so the viscosity of the fluid is 
\begin{equation}
    \eta = \eta_\infty + (\eta_0 - \eta_\infty) \left(1+ \lambda^2 \abs{\dot \gamma}^2 \right)^{(n-1)/2}
\end{equation}
with by default $\eta_0 = 9.437 \times 10^{-3} \si{.Pa. s}$, $ \eta_\infty=0.001 \si{.Pa. s}$, $\lambda = 0.6054 \si{.s}$  and $n = 0.5328$ corresponding to the XG solution with $c/c^* = 0.2$. 

\begin{figure*}[!ht]
\centering
\includegraphics[width=.7\textwidth]{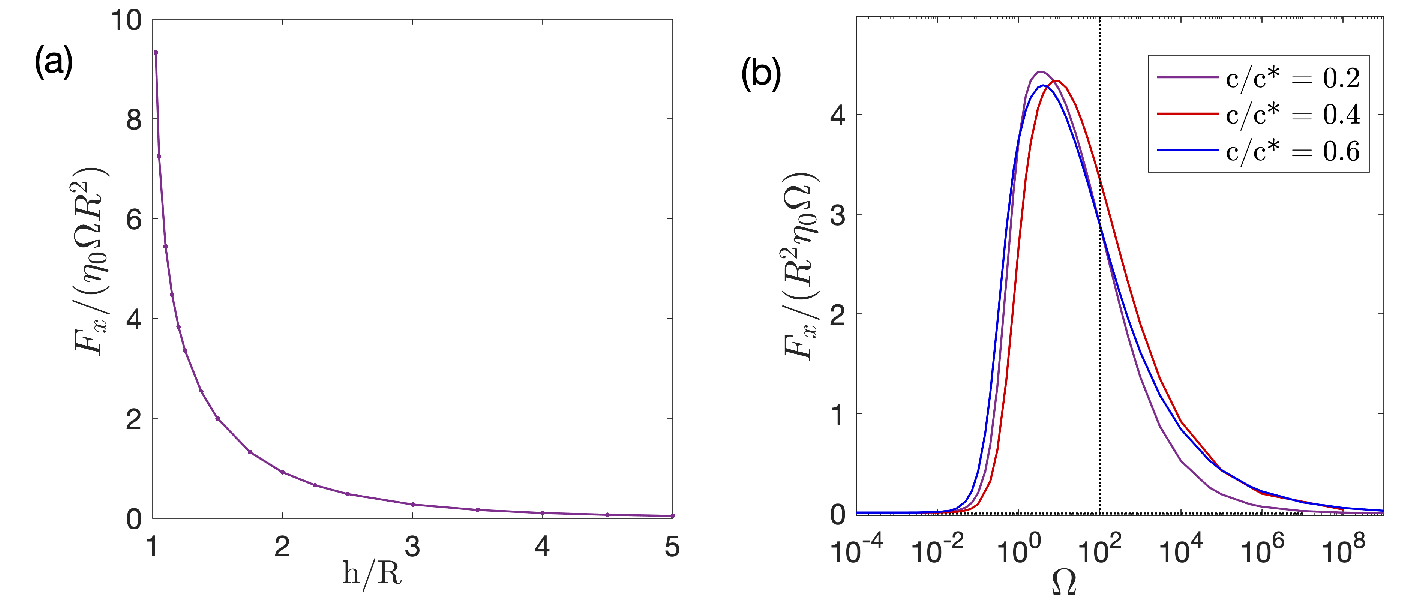}
\vspace{-2mm}
\caption{Rotation-induced drift force without an external flow (a) depending on the height of the cylinder $h$ and (b) on the fluid type.  }
\label{fig:SIwvdrift}
\end{figure*}

\subsection{Advection of a cylinder by a shear flow }
\begin{figure*}[!ht]
\centering
\includegraphics[width=.6\textwidth]{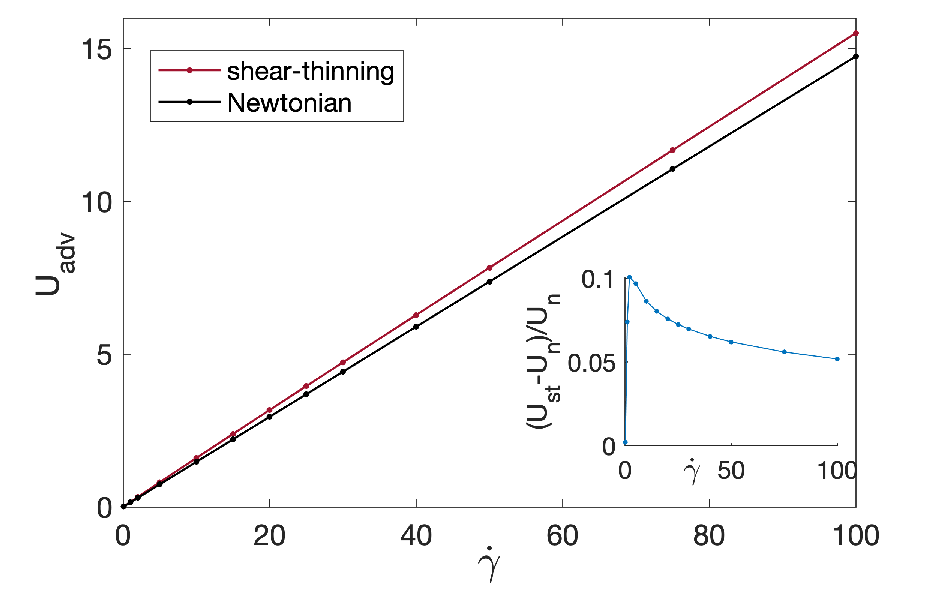}
\vspace{-2mm}
\caption{ Advection speed of a force-free cylinder above a wall in a Newtonian and a shear-thinning fluid. Inset: the relative difference between the two cases is below $10 \%$. }
\label{fig:SIwv}
\end{figure*}

To estimate how shear thinning affects the weathervane effect, we consider a non-rotating cylinder in a shear flow above a wall. In this case, we set an unknown translation speed $U_{adv}$ as the boundary condition. Solving the Stokes equations with a force-free condition yields the value of the induced translational velocity depending on $\dot \gamma$ for a Newtonian and shear-thinning fluid, as plotted in SM Fig~\ref{fig:SIwv}. We find that although non-linear effects occur, they are very weak and cannot account for the rheotactic abilities of the swimmers in a shear-thinning medium. 

\end{document}